\documentclass[aps,showpacs]{elsarticle}

\usepackage{amsmath,amssymb,graphicx,float,color}

\usepackage{amsmath}
\usepackage{xspace}
\usepackage{appendix}
\usepackage{float}
\usepackage{algorithm2e}
\usepackage{url}

\begin{document}

\title{An Empirically grounded Agent Based simulator for the Air Traffic Management in the SESAR scenario}


\author[westminster]{G. Gurtner\corref{correspondingauthor}}
\author[palermo]{C. Bongiorno}
\author[deepblue]{M. Ducci}
\author[palermo]{S. Miccich\`e}

\address[westminster]{Department of Transport and Planning, University of Westminster, 35 Marylebone Road, London NW1 5LS, United Kingdom}
\address[palermo]{Dipartimento di Fisica e Chimica, Universit\`a di Palermo, Viale delle Scienze, Ed. 18, I-90128, Palermo, Italy}
\address[deepblue]{Deep Blue srl, P.zza Buenos Aires, 20, I-00100, Roma, Italy}

\cortext[correspondingauthor]{Corresponding author}

%
%
%

%
%

\date{\today}

\begin{abstract}

In this paper we present a simulator allowing to perform policy experiments relative to the air traffic management. Different SESAR solutions can be implemented in the model to see the reaction of the different stakeholders as well as other relevant metrics (delays, safety, etc). The model describes both the strategic phase associated to the planning of the flight trajectories and the tactical modifications occurring in the en-route phase. An implementation of the model is available as open-source and freely accessible by any user.

More specifically, different procedures related to business trajectories and free-routing are tested and we illustrate the capabilities of the model on airspace implementing these concepts. After performing numerical simulations with the model, we show that in a free-routing scenario the controllers perform less operations although they are dispersed over a larger portion of the airspace. This can potentially increase the complexity of conflict detection and resolution for controllers.

In order to investigate this specific aspect, we consider some metrics used to measure traffic complexity. We first show that in non-free-routing situations our simulator deals with complexity in a way similar to what humans would do. This allows us to be confident that the results of our numerical simulations relative to the free-routing can reasonably forecast how human controllers would behave in this new situation. Specifically, our numerical simulations show that most of the complexity metrics decrease with free-routing, while the few metrics which increase are all linked to the flight level changes. This is a non-trivial result since intuitively the complexity should increase with free-routing because of problematic geometries and more dispersed conflicts over the airspace.

\end{abstract}

\maketitle

\section{Introduction}

In 2012 around 9,5 million flights crossed the European airspace and this number is expected to increase by 50\% in the next 20 years \cite{Eurocontrol2013}. Due to this traffic increase, without significant changes in the way air transport is currently managed, flying in Europe could lead to increased costs for the airlines, due to greater delays, and for the environment, due to higher CO$_2$ emissions. To tackle these challenges, the European Commission created in 2007 the SESAR JU (Single European Sky ATM Research Joint Undertaking) with the scope of coordinating all relevant research and development efforts in Europe. Since then, SESAR has been working on defining, exploring, testing, and implementing new solutions to cope with the foreseen air traffic increase. Among these, the concepts of free-routing and 4D trajectories have been proposed \cite{SESAR2007}\cite{EC2010}\cite{Eurocontrol2005} and are already implemented in some areas of the European Airspace \cite{EUROCONTROL2012}. In the future, or in what we call in the following, SESAR scenario, all airspace users will be allowed to plan an optimal trajectory, in space and time, from departure to arrival. At the moment, instead, aircraft need to stick to the structure of the airspace network and follow pre-defined airways which are often not the most direct routes. In this environment, air traffic controllers have the role of avoiding conflicts in some specific areas that are mainly located where airways intersect. The implementation of free-routing poses therefore some challenges in terms of safety of the operations and of complexity of the situation that controllers have to manage. For instance, conflicts may become harder to detect due to the spread and increased number of possible conflicting points. Moreover, methods used to solve conflicts (i.e. direct routes) may not be applicable any more since aircraft will be already flying the most direct trajectory.

Although Free Route Airspace is already implemented in some parts of Europe \cite{free-routing}, its application is still limited to conditions where traffic load is quite low. Therefore, it is relevant to assess its impact in the higher traffic conditions foreseen in the next 20 years in relation to the safety of the operations and to the complexity that controllers will have to manage. To this end, we have developed an Air Traffic Simulator to evaluate the implementation of some of the features foreseen by SESAR. The Simulator consists of several modules that can work independently to allow the user to perform analyses at different levels, from the creation of the airspace to the definition of the planned trajectories up to the execution phase. The Simulator takes input both real data (actual trajectories, airspace structures) or to use fully synthetic data, and it can thus simulate both current and future scenarios and all the conditions in between. Even though other similar tools exists (FACET \cite{tumer-agogino_aamas07}, CATS \cite{Allignol2011,Allignol2013}), we developed our own simulator to have an open-source tool intended as (i) something flexible to perform analyses in a quick and easy way considering only relevant parameters and (ii) providing a library of tools for investigating the ATM system.   

In this paper we present our simulator and the results obtained on the analysis of the implementation of Free Route from a safety and a complexity perspective. In particular we show that by progressively making trajectories straighter, the number of actions that controllers have to perform to ensure aircraft safety, i.e. to avoid conflicts, are decreasing. But the area where these actions have to take place is more scattered thus potentially increasing the complexity of the situation  controllers have to manage. Given that there is no consensus, as far as we know, about what is a complex situation for controllers, we first compared some complexity metrics that can be found in literature \cite{Histon2002} with our own (i.e. the number of controller's actions) and found out that they are well-correlated. In addition we show that our own complexity metrics are very well related with what a human would be experiencing in terms of complexity \cite{Sridhar1998,Delahaye2000,Chatterji2001,Laudeman1998}. 

The paper is organized as follows. Section \ref{datasec} described the data we used as input to calibrate the model in the current scenario. We also show how the complexity metrics change in the SESAR scenario and what are the main factors contributing to the decrease of the overall complexity that we measured. Section \ref{modeltext} is describing the Simulator, with a focus on the modules that have been used in the analysis, while Section \ref{validation} presents the results of the calibration process. Section \ref{results} shows the results of the analysis when implementing free-routing within the simulator. We finally present some conclusions in section \ref{concl}.

\section{Data} \label{datasec}

The database we use contains information on all the flights that, even partly, cross the ECAC airspace. Data are collected by EUROCONTROL (\url{http://www.eurocontrol.int}), the European public institution that coordinates and plans air traffic control for all Europe and were obtained as part of the {\em{SESAR Joint Undertaking}} WP-E  research project ELSA ``Empirically grounded agent based model for the future ATM scenario'' \footnote{Data can be accessed by asking permission to the legitimate owner (EUROCONTROL). The owners reserve the right to grant/deny access to data.}.

Data come from two different sources. First, we have access to the Demand Data Repository (DDR) \cite{DDR} database containing all the trajectories followed by any aircraft in the ECAC airspace. Indeed, we have access to data relative to a time period of 15 months, from the $8^{th}$ of April 2010 to the $27^{th}$ of June 2011. Each 28 day time period is termed AIRAC cycle. A planned or realized trajectory is made by  a sequence of navigation points crossed by the aircraft, together with altitudes and timestamps. The typical time between two navigation points lies between 1 and 10 minutes, giving a good time resolution for trajectories. In this paper we use the ``last filed flight plans'', i.e. the so-called M1 files, which are the planned trajectories -- filed from 6 months to one or two hours before the real departure. We also use the real trajectories, i.e. the so-called M3 files, because we will compare planned and realized trajectories in order to investigate the role of the air traffic controllers. 

The database includes all flights in the enlarged ECAC airspace\footnote{Countries in the enlarged ECAC space are: Iceland (BI), Kosovo (BK), Belgium (EB), Germany-civil (ED), Estonia (EE), Finland (EF), UK (EG), Netherlands (EH), Ireland (EI), Denmark (EK), Luxembourg (EL), Norway (EN), Poland (EP), Sweden (ES), Germany-military (ET), Latvia (EV), Lithuania (EY), Albania (LA), Bulgaria (LB), Cyprus (LC), Croatia (LD), Spain (LE), France (LF), Greece (LG), Hungary (LH), Italy (LI), Slovenia (LJ), Czech Republic (LK), Malta (LM), Monaco (LN), Austria (LO), Portugal (LP), Bosnia-Herzegovina (LQ), Romania (LR), Switzerland (LS), Turkey (LT), Moldova (LU), Macedonia (LW), Gibraltar (LX), Serbia-Montenegro (LY), Slovakia (LZ), Armenia (UD), Georgia (UG), Ukraine (UK).} even if they departed and/or landed in airports external to the enlarged ECAC airspace.

The other source of information is given by the NEVAC files. NEVAC files \cite{NEVAC} contain the definition (borders, altitude, relationships, time of opening and closing) of airspace elements, namely airblocks, sectors, FIR, etc. The active elements at a given time constitute the configuration of the airspace at that time. Thus, they give the configuration of the airspaces for an entire AIRAC cycle. Here we only use the information on sectors, FIRs or ACCs and configurations to rebuild the European airspace. Specifically, at each time we have the full three dimensional boundaries of each individual sector and FIR or ACC in Europe. All this information has been gathered in a unique database, using MySQL, in order to allow fast crossed queries. 

In this study we considered the LIRR ACC (Rome, Italy) between 2010-05-06 and 2010-06-03, i.e. the 334 AIRAC. We are considering only commercial flights, which in first approximation is obtained by considering 1) flights performed with Landplanes (i.e. no helicopter, gyrocopter, only aircraft which can only operate from or alight on land), 2) scheduled flights, 3) flights with a IATA code 4) flights with a duration longer than 10 minutes. We also excluded all flights having a duration shorter than 10 minutes and a few other flights having obvious recording data errors. To focus our attention on the en-route phase we filtered out from the flight plans all navpoints crossed at an altitude lower then FL 240. After the filtering procedure, 35714 flights were retained in the entire AIRAC. 

In order to include the local constraints of the sector capacities, it is important to remember that the sectors are not static geometric regions but they are merged together and splitted dynamically  to fullfil workload requirements. For the sake of simplicity we will refer to the collapsed sector defined in the reference \cite{Gurtner2015}. These are a static bi-dimensional projection of the sectors higher than FL 350. The sectors capacity inferred from data is defined as the maximum number of flights expected within a time-window of one hour inside the collapsed sector. Finally, data do not include Saturdays and Sundays in order to avoid weekly seasonality effects.


\section{The Model} \label{modeltext}

The ELSA Air Traffic Simulator is composed of several fairly independent modules:
\begin{itemize}
\item A \textbf{network generator}, including navigation points and sectors.
\item A \textbf{strategic layer}, able to investigate the competition amongst airlines for the strategic allocation of best routes and here used as a \textbf{flight plan generator}.
\item A \textbf{rectification module}, used to straighten up trajectories when simulating free-routing.
\item A \textbf{pre-tactical de-conflicting module}, used to generate ({\em{by brute-force}}) conflict-free planned trajectories starting from real or surrogate planned trajectories.
\item A \textbf{tactical layer}, with a conflict resolution engine, simulating a tunable, imperfect super-controller.
\item A \textbf{post-processing} module, including standard metrics computation and a simple graphic interface to see the networks computed from trajectories produced by the model.
\end{itemize}

\begin{figure}[H]
\centering
                 \includegraphics[width=0.65\textwidth]{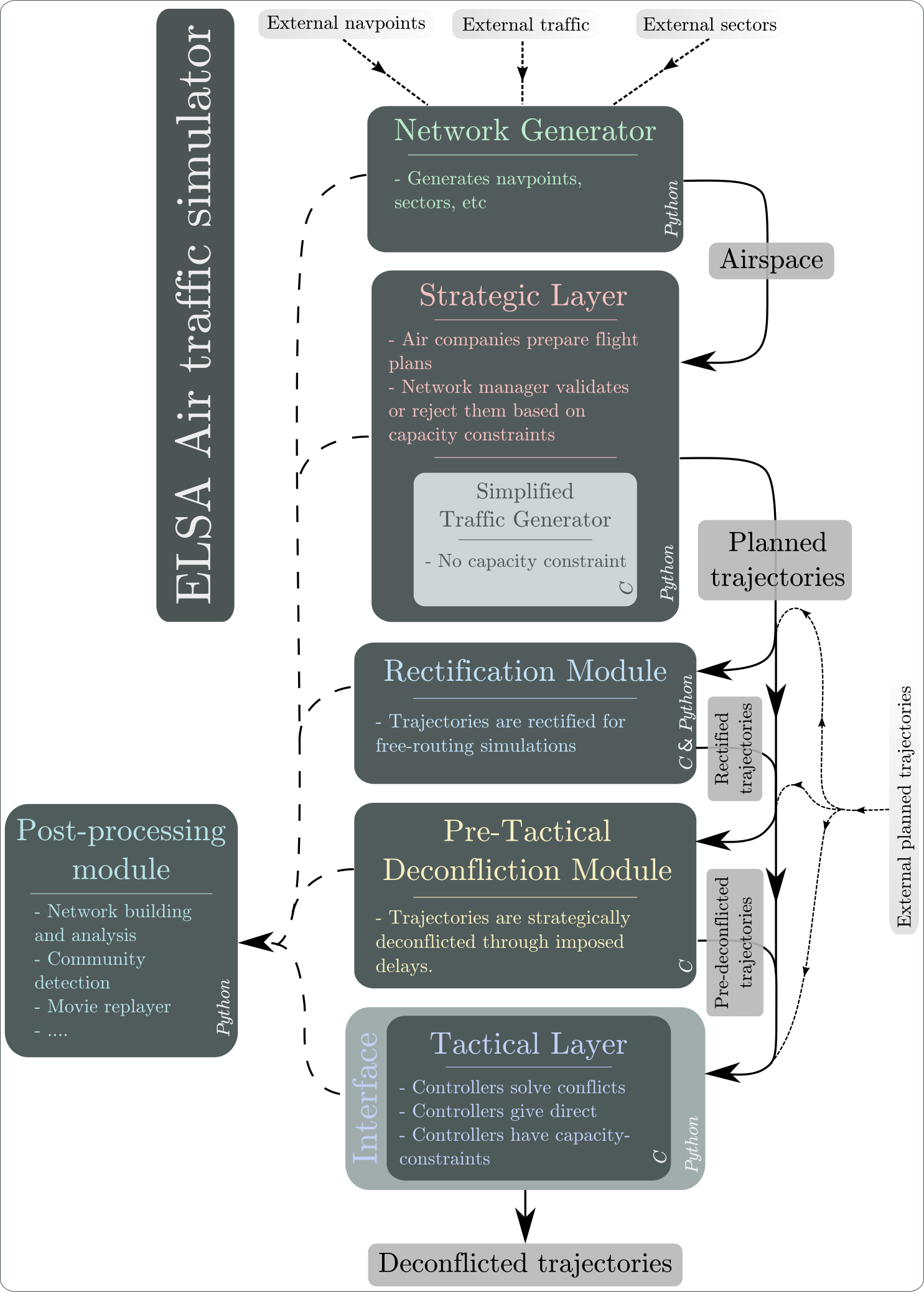}
\caption{Organization of the model.}
\label{fig:modeltext}
\end{figure}

A schematic representation of the code is displayed in Fig. \ref{fig:modeltext}.
The model uses a description of the airspace in terms of navigation points (which form a network on which the flights are travelling) and sectors. The latter defines in which area the controller can actually interact with the flights.

The model includes a possibility of building some airspace, ranging from full manual definitions of navigation points and sectors to automatic generation of airspace. Real airspaces can also be used easily, possibly integrating deviations from reality controlled by the user (e.g. number of navigation points). The airspace can be composed of several sectors and is in three dimensions.

The model gives the possibility of generating flight trajectories in a controlled way. In particular, there exists the possibility of slowly tending to business trajectories -- i.e. straighter and straighter trajectories, thus testing the possibility of continuous integration of the free-route scenario on real airspace. The integration can be heterogeneous, with some sectors keeping a fixed grid of navigation points whereas others have already moved to business trajectories.

The core engine of the model is the resolution of conflicts, given an airspace and some flights traveling through it. The controller can have different strategies (horizontal deviation, vertical deviations, give a direct) depending on the environment and its own limited -- and tunable -- look-ahead time. The controller is perfect in the sense that it avoids all conflicts but sometimes makes suboptimal decisions due to its limited forecast capacities. ``Shocks'' can be used to perturb the trajectories. Indeed, some parts of the airspace can be randomly shut down for a given time, simulating weather events or military exercises. The spatial and temporal distributions of shocks can be fully controlled.


The simulator code is written in Python and C \cite{Kernighan1988}, but only a limited knowledge of Python and C is required in order to use it. It is released under the General Public License version 3, i.e. it is open-source. In particular, it is freely downloadable on Github at the address \url{https://github.com/ELSA-project/ELSA-ABM}. The community is welcome to use it, modify it, ask for clarifications and report bugs, using the tools available on GitHub or contacting directly the authors. The code has currently been fully tested under Linux, but should work also with MACOS with minimal effort.

Each of the modules briefly illustrated below are fully described in \cite{NEVAC}.

\subsection{Network Generator} \label{NetGenModule}

The starting point of our simulator is a network generator module that is used to generate the navigation point grid that will then be populated by aircraft. In particular, it allows:
\begin{itemize}
\item To generate the spatial distribution of navigation points or use external data,
\item To compute the navigation points network edges with a triangulation \footnote{We use the Delaunay triangulation for its properties \cite{Delaunay1934}.} or use external data,
\item To generate sectors at random, using a Voronoi tessellation \cite{Voronoi1908} for the boundaries or use external data,
\item To compute time of travels between edges of navigation points or use external data.
\end{itemize}
Hence the user can fully specify the network and the sectors or use the module in a semi-automated way. It is also possible to build a network based on traffic data.

\subsection{Strategic Layer} \label{StratLayerModule}

The strategic layer is a full Agent-Based Model where different agents, i.e. airlines and network manager, are collaborating or competing for the same resources, i.e. time slots and trajectories. An earlier version has been described in detail in \cite{Affronti2014,Gurtner2015}. The strategic layer is designed to generate trajectories with a coarse level of description, suitable to study high level phenomena. In particular, the trajectories are kinematic and do not take into account winds, weight, etc.

Some results related to this model have already been presented in Ref.  \cite{Gurtner2015}. With respect to that work, we updated the model in order to have a full, navigation point-based description of the airspace, on top of the sector-based description. The full description of this updated version is available in \cite{git_repo_model_description}. In short, the strategic layer takes as input a network of navigation points with the legitimate paths and fills the airspace realistically, with airlines submitting flight plans and the network manager rejecting or accepting them on a capacity-constraint basis. 

For the purposes of the present work, we consider the strategic layer of our simulator as a tool that produces a set of realistic trajectories, fulfilling some capacity constraints, which can be used as an input to the tactical layer below. Since the agents can have different behaviors, the results depend also on the choices of the code user. Here we used some default values that were obtained by considering the results of the calibration procedure we ran in \cite{Affronti2014}.


As a result, the strategic layer, when used as a  ``traffic generator'', generates synthetic traffic on a given network of navigation points and sectors. It allows to create traffic in a set of sectors given some airports and/or entry/exit points in a realistic way, making sure no sector is overloaded. The user can specify in particular:
\begin{itemize}
\item A total number of flights,
\item the navigation point network to be used,
\item a distribution of flights per pair of entry/exit points,
\item a distribution of departure times,
\item some capacities for the sectors.
\end{itemize}
Other parameters, such as those specifying the type of airlines present in the considered airspace, are set to default values, see Table \ref{StratPar}.


We also have a ``simplified traffic generator'' that simply generates trajectories by randomly extracting navigation points according to the constraints mentioned previously and without considering the sectors capacities. We used this simplified approach because it performs faster than the full traffic generator in which the capacity is set to infinity. Having a fast simulator for trajectories that do not fulfill capacity constraints revealed itself useful when performing numerical simulations for the description of the SESAR scenario.

It is worth mentioning that both traffic generators give as an output 2-D trajectories, i.e. trajectories that lie on an horizontal plane. In order to have 3-D trajectories we implemented the following procedure: (i) we first extract from the distribution of flight levels occupancy a number of flight level values equal to the number of navigation points of the considered trajectory; (ii) we then order the values so that the first third of them are in increasing order, the last third of values are in decreasing order the second third of values are mixed. This is a very simplistic procedure that nevertheless has the advantage to roughly capture the fact that aircraft have an ascending en-route and descending phase.

When generating the planned trajectories, we consider that all aircraft travel with the same velocity as a first approximation. This velocity is taken equal to the average velocity of all aircraft present in the considered
airspace. However, trajectories can also be taken from real data, in which case we consider the real velocity of each trajectory segment. 

Given the previous considerations, when generating the flight plan trajectories, one has also to consider the following additional constraints:
\begin{itemize}
\item a distribution of flight levels occupancy,
\item a distribution of velocities.
\end{itemize}

\subsection{Rectification Module} \label{RectModule}

The rectification module was introduced to study the transition between the current scenario and the SESAR scenario in a controlled way. In particular, there exists the possibility of slowly tending to straight trajectories, thus testing the possibility of progressive integration of the free-route concept on the current airspace structure. The integration can be heterogeneous, with some sectors keeping a fixed grid of navigation points whereas others have already moved to free-route. The module requires as input a generic M1 file, i.e. a set of planned trajectories in the current scenario, and produces as output another M1 file where trajectories have larger target value of efficiency. This metric is defined as the ratio between the actual length of the trajectory and the shortest path between origin and destination \cite{Bongiorno2015a}:
\begin{eqnarray}
                           E =\frac{ \sum_{N_f} d(O,D)_i }{\sum_{N_f} d_{BP}(O,D)}
\end{eqnarray}
At each step the algorithm evaluates the current Efficiency and if it is less than the target Efficiency it substitutes a point of a route randomly selected with the medium point between the previous navigation point and the successive.

\subsection{Pre-Tactical De-conflicting Module} \label{PreTactModule}

The model also includes a pre-tactical de-conflicting module. Its task is to generate conflict-free planned trajectories starting from real or surrogate planned trajectories. The need for such a module is due to the fact that one of the features foreseen by SESAR will be a better planning of the trajectories such that they may be already conflict-free \cite{EUROCONTROL2012a}. As such, since we are still at the planning level and no issue regarding the flight conditions is taken in consideration, differently from the module of \ref{TactLayerModule}, that modifies the sequence of navigation points/flight levels, this module only acts on the departure time of the aircraft.

\subsection{Tactical Layer} \label{TactLayerModule}

The tactical layer of our simulator can be considered as a zero-intelligence agent-based model (in the line of \cite{Farmer2005}) that simulates the interactions between ATC controllers and aircraft where we do not have learning and the agents interact in a mechanistic way.

The core of the tactical layer is composed of the ``conflict detection'' and ``conflict resolution'' submodules \footnote{The issue of conflict detection and resolution has been extensively studied in the literature. We can refer the reader to a short review, available at {\tt{http://www.complexworld.eu/agent-based-models-take-off/}}}. This module is used to check whether any conflict occurs and to solve it \cite{Bongiorno2015a}. It essentially works along the same lines than the analogous module already presented in Ref. \cite{Bongiorno2013}. In order to check for collisions between the $i$-th aircraft and all the other $N_f-1$ ones, the Conflict Detection module  performs a sample of the aircraft trajectories at $\delta t$ seconds. This time-step is then divided into N elementary time increments $\delta t$ and the position of each aircraft is computed at each of them. We thus have an array of positions for each aircraft simultaneously present in the considered airspace. Then the module compares the position-array of the i-aircraft with the position-arrays of the other $N_f-1$ aircraft by calculating the distance between any two aircraft at each time increment. If at least one value is below the minimum separation distance of 5 nautical miles then a conflict is detected.

The simplest version of the Conflict Resolution module is based on two strategies: rerouting and flight level change. In order to solve the conflicts the module first tries to reroute the aircraft then to change its flight level if the rerouting fails. To reroute the aircraft the Simulator generates a set of random temporary navigation points around the location of the possible conflict within a range of 100 km. Then it tries to send the aircraft towards one of these temporary navigation points selected with the criteria of (i) minimizing the total path length and (ii) with the constraint that the angle between planned and deviated trajectories must be smaller than a threshold value selected by the user. If no solution is found by performing a re-routing, then the module tries to change the flight level. It first tries to send the aircraft 2 FL up then it tries 2 FL down, if it fails.

\subsection{Post Processing} \label{PostProcModule}

The simulator also features some post-processing tools which help the user having a better grasp on the results of the model. The functions currently include for instance network-builders from trajectories, computation of different network metrics, comparisons of networks, as well as a movie maker to replay simulations.

\section{Calibration and Data Input} \label{validation}

In this section we summarize the parameters entering the model and the  activities performed for the selection of the parameters that were calibrated, when possible, on real data. We gather all parameters of the simulator into two main categories. On one side we have the parameters needed for the generation of the flight trajectories, either synthetic or taken from real data. These parameters mainly refer to the parts of the model described in section \ref{NetGenModule}, \ref{StratLayerModule}. On the other side we have the set of parameters needed for the management of the trajectories. These parameters mainly refer to the parts of the model described in section \ref{TactLayerModule}.

\subsection{Trajectory Generation}\label{sec:validation}

The main operational inputs for the strategic layer have been collected during interviews with Alitalia Flight Dispatchers that work at the Alitalia Operation Center (OCC)\footnote{This small section is taken directly from the SID paper 2013 \cite{Bongiorno2013}.}. They are the professional figures in charge of defining the flight plans and monitoring the flight execution phase. The Alitalia Operation Center is responsible of coordinating and managing almost 700 flight per day, of which around 70 are long-haul flights. 
The information collected that are more relevant for the development of the strategic ABM are mainly related to: 
$(a)$ the timeframe of the flight plan definition process (6 hours in advance for long-haul flights and 2-hours in advance for medium and short-haul flights);
$(b)$ the costs taken into account for the flight plan optimization;
$(c)$ the interactions between the Flight Dispatcher and the Network Manager and the fact that the flight dispatcher is unaware of other companies strategies;
$(d)$ the flight plan submission process, how flight plans are rejected and submitted again for the final approval;
$(e)$ the criticalities related to the planning phase such as  the  exceeding of capacity of one or more sector, bad weather avoidance, partial or total closure of destination airport or unpredictable events like strikes, big events, wars.

In Table \ref{StratPar} we list all the parameters relevant for the generation of flight trajectories. In the third column we give a short description of the parameters and in the fourth column we introduce a classification of the parameters in terms of the three categories described below:
\begin{itemize}
\item FP - free parameter, to be chosen at will depending on the type of experiments one wants to perform.
\item EO - external output, to be chosen according to some rule depending on the type of realistic experiments one wants to perform.
\item CD - parameter that needs to be calibrated from data.
\item CV - parameter that needs to be calibrated according to the validation activities performed with ATM experts and ATCOs.
\end{itemize}
Depending on the way the different modules are used, the same parameter can fit into different categories, corresponding to different calibration procedures.

\begin{table}
\caption{Model parameters relevant for the generation of flight trajectories.  In boldface we indicate the category associated to the parameter in the present work.}  \label{StratPar}
\centering
\begin{tabular}{|c|c|c|c|c|}
\hline
{\bf{ID}} & {\bf{Parameter}}  & {\bf{Description}}  			& {\bf{Type}} 				& {\bf{Value}}\\
\hline
01 & $N_s$ & Number of sectors									& FP/\textbf{EO} 				& 10\\ 
\hline
02 & $N_p$ & Number of navigation points per sector  			& FP/\textbf{EO} 				& \begin{tabular}{c}Dis. from data\\($\sim 28$ in average)\end{tabular} \\ 
\hline
03 & $\mathcal{A}$ & Area of the considered airspace  			& FP/\textbf{EO} 				& LIRR \\
\hline
04 & $N_a$ & Number of airports or entries/exits  				& FP/\textbf{EO}  			& 189 \\ 
\hline
05 & $\{t_{ij}\}$ & Crossing times of edges  					&\textbf{CD}					& Dis. from data \\
\hline
06 & $\{\mathcal{C}_\alpha\}$ & Capacity of airports 			& FP/CD			  			& Not used here \\
\hline
07 & $\{\mathcal{C}_i\}$ & Capacity of sectors  				& FP/{\bf{CD}} 				& Dis. from data \\ \hline
08 & $d_{min}$ & 
		\begin{tabular}{c}
			Minimum number of navpoints between \\
			entry and exit potentially linked by a flight
		\end{tabular}  
													& FP/{\bf{CD}} 				& 5 \\ 
\hline
9 & $N_{fp}$ & Number of flight plans submitted for each flight 	& FP/{\bf{CV}} 				& 10 \\ 
\hline
10 & $N_{sp}$ & 
		\begin{tabular}{c}
			Number of paths of navigation points\\
			per path of sector 
		\end{tabular}
													& {\bf{CD}}				& 2 \\ 
\hline
11 & $\tau$ & Time shifting step for the flight plan  				& FP/{\bf{CV}} 				& 15 min.\\ 
\hline
12 & $DP$ & Type of departure times pattern		    			& FP/{\bf{CD}}/EO			& Dis. from data\\ 
\hline
13 & $\delta t_0$ & 
		\begin{tabular}{c}
			Standard deviation of noise added \\
			to the real departure times
		\end{tabular} 
													& FP				& Not used here \\
\hline
14 & $N_f$ & Number of flights 							& {\bf{FP}}/EO 				& Variable \\ 
\hline
15 & $\alpha$, $\beta$ & Behavioural parameters for airlines 		& FP/{\bf{CD}} 				& 1, 0.001\\ 
\hline
16 & $N_{shock}$ & Number of sectors which are shut down 		& FP	 			& Not used here\\ 
\hline
\end{tabular}
\end{table}




In addition to these parameters, the flight plan generator requires the specification of the distributions indicated in Table \ref{modinput}. 
\begin{table}[H]
\caption{Distributions relevant for the generation of flight trajectories.}  \label{modinput}
\centering
\begin{tabular}{|c|c|l|}
\hline
{\bf{ID}} & {\bf{Param.}}  & {\bf{Description}}  \\
\hline
1 & $v$ &  distribution of the aircraft velocity  \\
\hline
2 & $FL$ & distribution of the flight levels occupancy \\
\hline
3 & $OP$ & distribution of flights between origin-destination pairs \\
\hline
4 & $DEP$ & distribution of departure times \\
\hline
\end{tabular}
\end{table}

\begin{figure} [H]
\centering
        {\includegraphics[width=7.0cm]{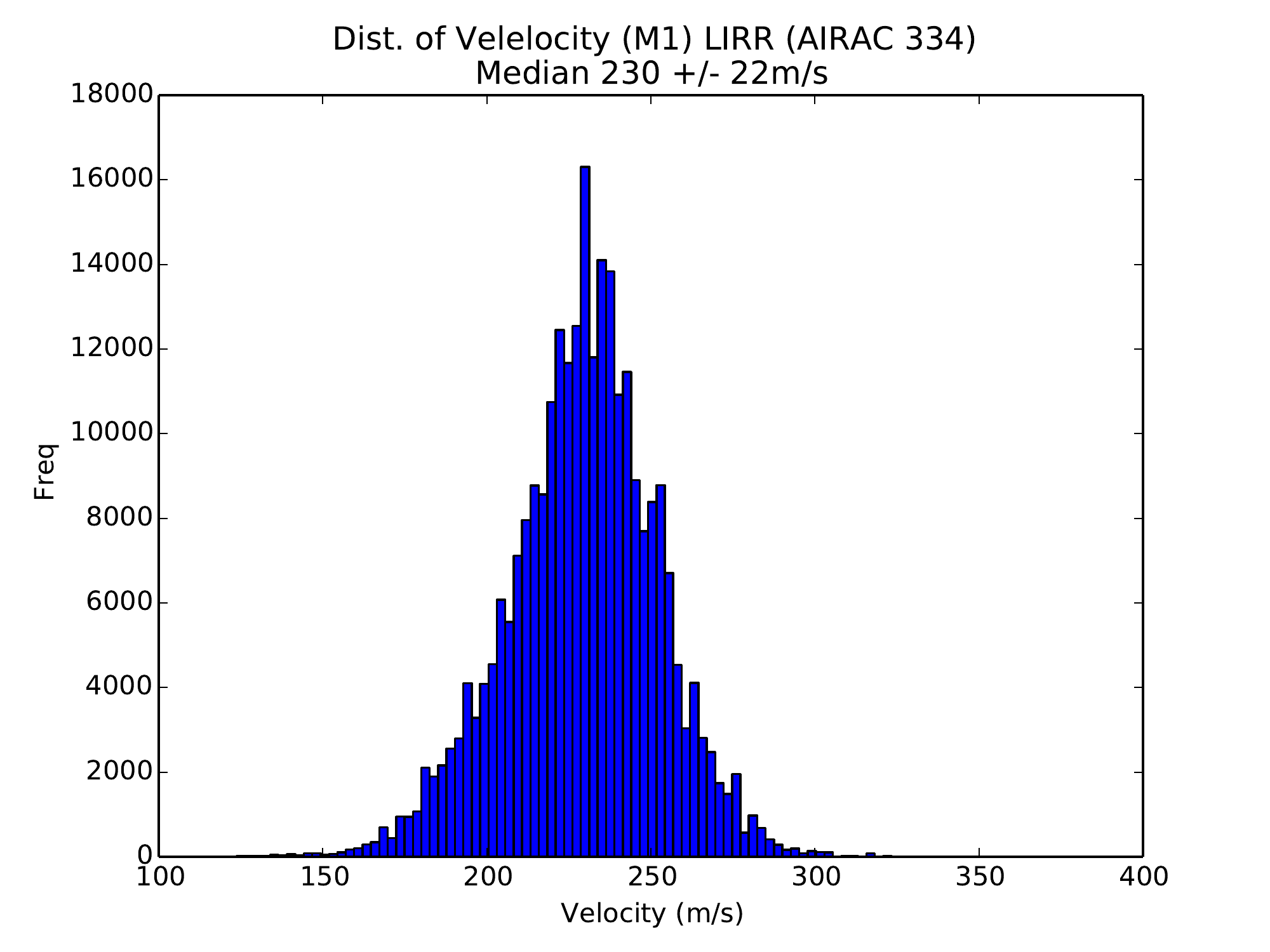}
         \includegraphics[width=7.0cm]{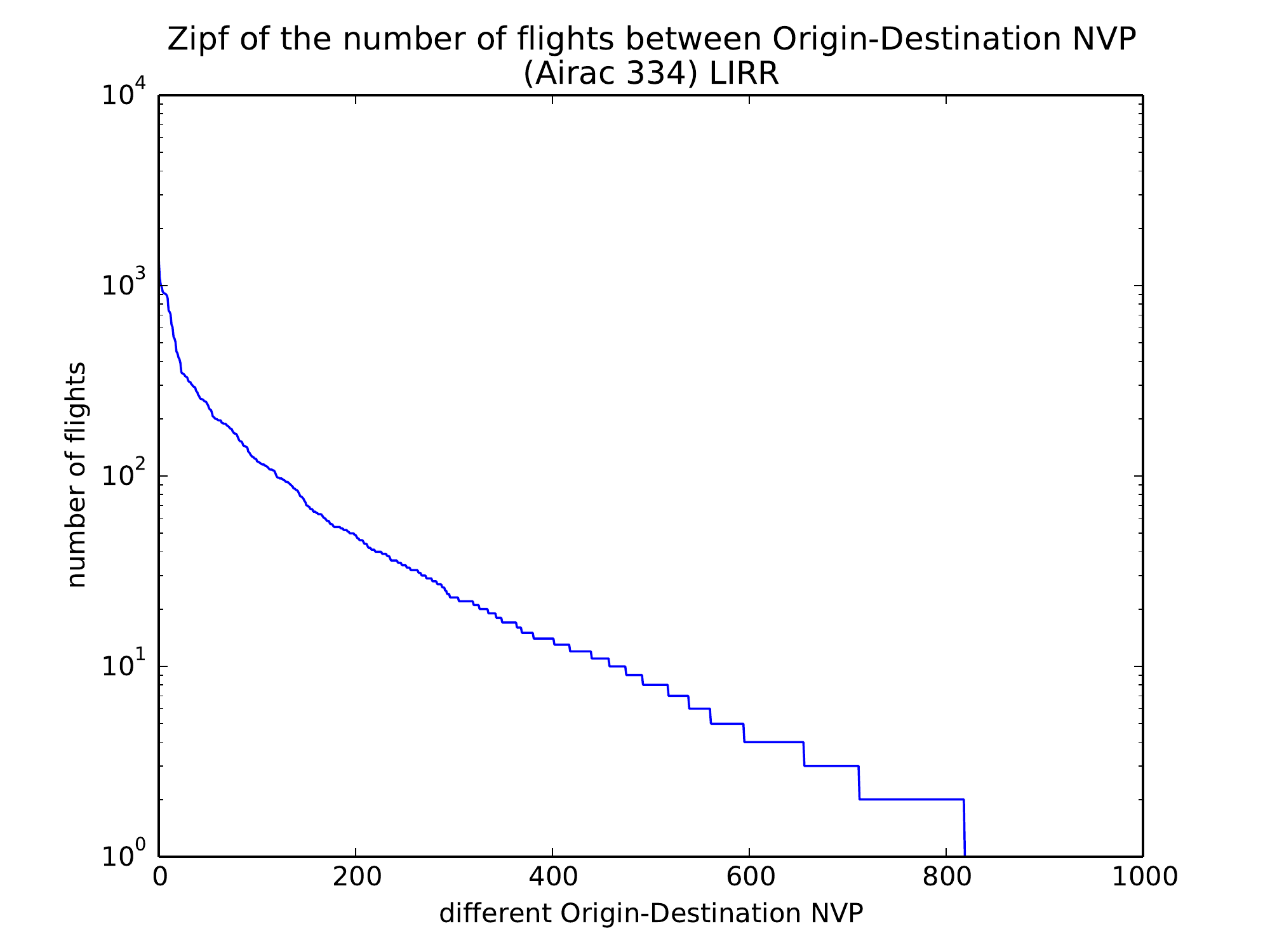}} \\
        {\includegraphics[width=7.0cm]{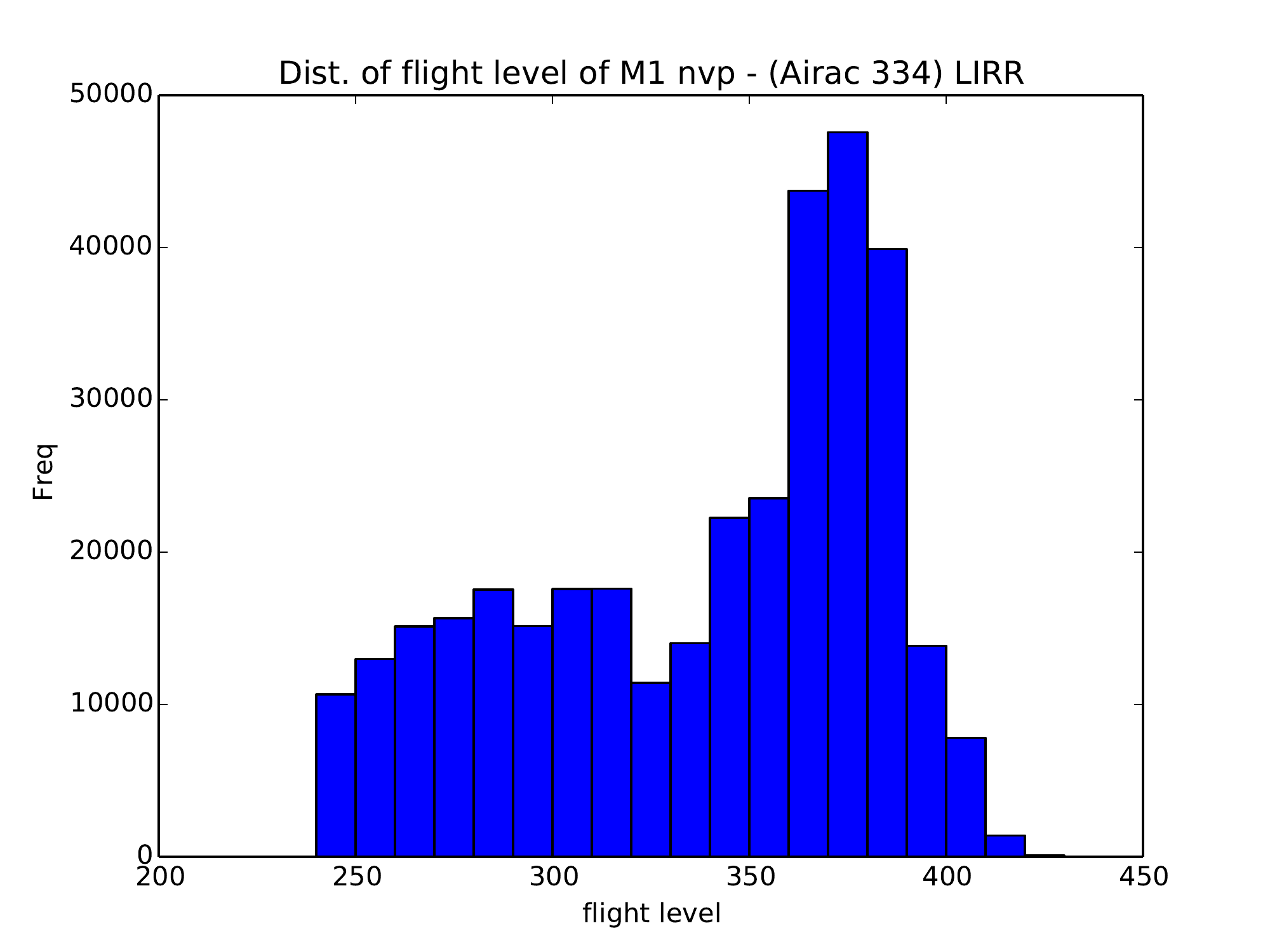}
         \includegraphics[width=7.0cm]{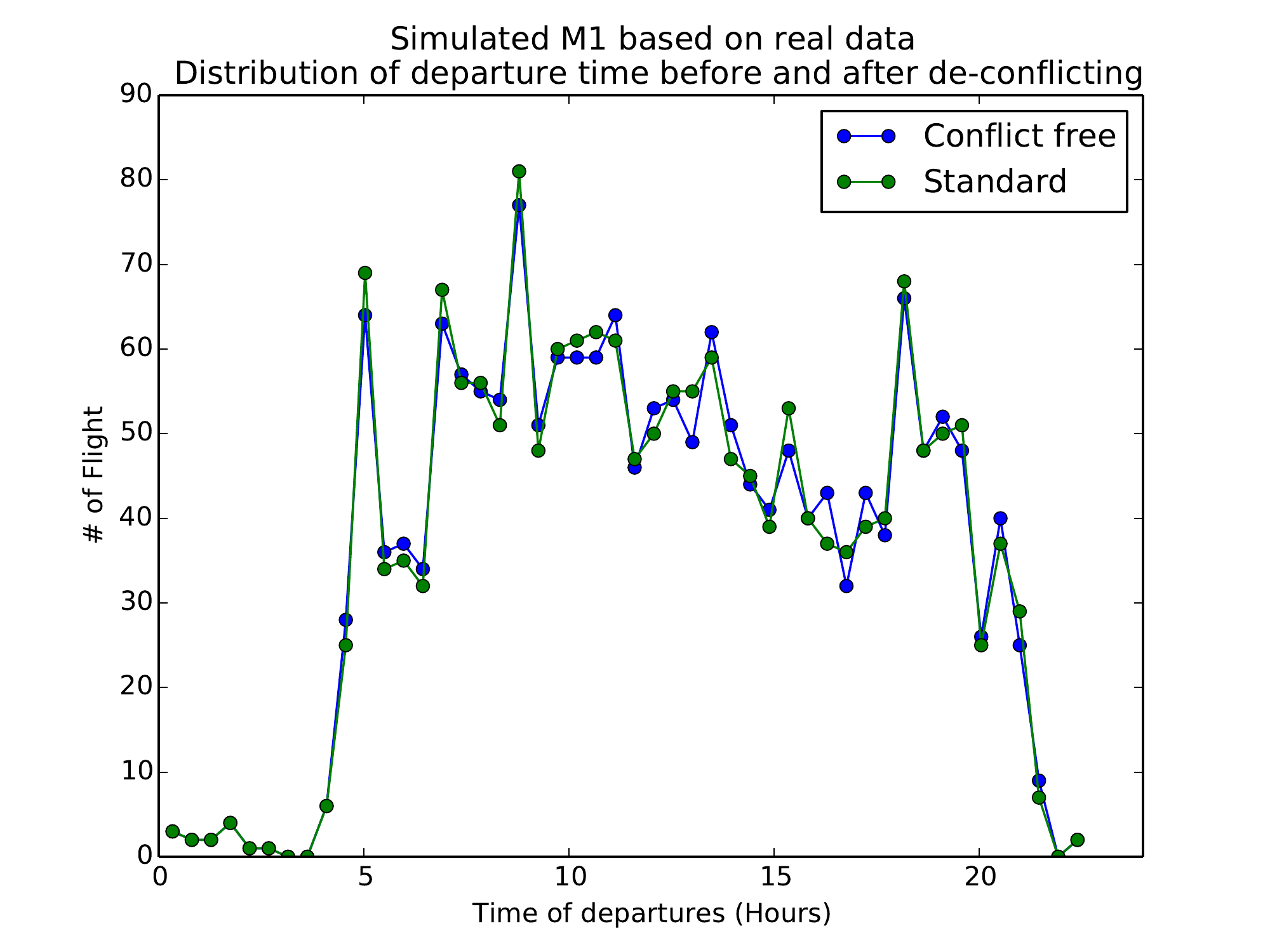} }
        \caption{Top-Left panel: velocity distribution measured from the M1 files. Top-Right panel: empirical probability distribution of the number of flights between origin and destination airports. Bottom-Left panel: empirical probability distribution of the flight level occupancy. Bottom-Right panel: distribution of time departure for the de-conflicted (blue line) and original M1 flight trajectories (green line). Data refer to the LIRR ACC in day 06/05/2010.}        \label{velo}
\end{figure}

These are distributions that can be easily obtained from real data. The main calibration activity conducted on real data regards the choice of the aircraft velocity. In the top-left panel of Fig. \ref{velo} we show the distribution of the aircraft velocity measure starting from the M1 files and in the LIRR ACC. The median of the distribution is $v=230$ m/s which correspond to $v=828$ km/h. This is the value that we used in the simulations of section \ref{results}. From this distribution one might estimate $v_{min} \approx 130$ m/s and $v_{max} \approx 320$  m/s. The standard deviation is about $\approx 22$ m/s. The top-right panel shows the distribution of flights between airport pairs, while the bottom left panel shows the distribution of the flight levels occupancy. Finally, in the bottom-right panel we show the distribution of the time departure for the original M1 flight trajectories (green line), i.e. the distribution required in section \ref{PreTactModule}. We also show the distribution in the case when the pre-tactical de-conflicting module of section \ref{PreTactModule} is applied. Interestingly, our procedure does not alter the original distribution in a significant way, notwithstanding the fact that it is a brute-force method. This is very important for us, because we can simulate a strategic conflict-free scenario with very realistic departure times. It is also interesting per se, because it means that real strategic deconfliction procedures could be implemented without disrupting the business models of the airlines.

\subsection{Trajectory Management}\label{sec:validation2}

The parameters related to the trajectory management are summarized in Table \ref{modpar}. 
The parameters that need to be calibrated from data are really few. However, there are many parameters (CV category) that are related to the behavior of controllers. In principle, these are parameters that could be inferred from data through some sophisticated data mining. However, we believe that these are the typical parameters that should be selected by consulting ATM experts and ATCOs. On the other hand, these are the parameters that one should change at will when performing scenario simulations to test how changing a certain feature will affect the ATM system.

\begin{table}[h]
\caption{Model parameters relevant for the management of flight trajectories.}  \label{modpar}
\centering
\begin{tabular}{|c|c|c|c|c|}
\hline
{\bf{ID}} & {\bf{Parameter}}  & {\bf{Description}}  & {\bf{Type}} & {\bf{Value}}\\
\hline
01 & $\Delta t$ &  \begin{tabular}{c}Length of the time-step.\\ 
									This is also related to the controller's look-ahead.
					\end{tabular}  																& FP & 12.27m \\ 
\hline
02 & $\delta t$ & Length of the elementary time-intervals.  									& FP & 8s \\
\hline
03 & $t_r$ & \begin{tabular}{c}Fraction of $\Delta t$ by which we move the overlapping\\
								 time-steps.
			\end{tabular}																		& FP & 0.25\\
\hline
04 & $\sigma_v$ &  \begin{tabular}{c}Range of the noise introduced in the estimation\\ 
									of the aircraft velocity
					\end{tabular} 																& CV & not used here\\
\hline
05 & $D_{max}$ & \begin{tabular}{c}Radius of the circle centered in B where we look\\
									 for temporary navigation points potentially relevant\\
									  for performing a re-routing.
				\end{tabular}																	& FP  & not used here \\
\hline
06 & $\alpha_M$ &  \begin{tabular}{c}Maximum angle of deviation between planned and \\
							modified trajectory.
					\end{tabular}																& CV  & 15 deg \\
\hline
07 & $T_{max}$ &  \begin{tabular}{c}Maximal temporal distance between the navigation\\
									 point B and navigation point E that identify when\\
									 a deviated portion of flight trajectory starts and ends
				\end{tabular}   																& FP/CV & 24.5m  \\
\hline
08 & $p_d$ & Probability to try to issue a direct. 											& CD/CV & 0\\
\hline
09 & $L_s$ & Sensitivity threshold for issuing a direct.  										& FP/CV & not used here\\
\hline
10 & $C_S$ & Center of each shock.  															& FP/CD & not used here\\
\hline
11 & $S_m$ & Average number of shocks per time-step per flight-level.  						& FP/CD & not used here \\
\hline
12 & $D_S$ & Temporal duration of each shock. 													& FP/CD  &  not used here\\
\hline
13 & $R_S$ & Radius of each shock.  															& FP/CD &  not used here\\
\hline
14 & $FL_{min/max}$ & \begin{tabular}{c}Minimum/maximum flight level where shocks\\
										 are generated
						\end{tabular} 															& FP/CD & not used here \\
\hline
\end{tabular}
\end{table}

\medskip 


\section{Results} \label{results}

In this section we show that our model predicts that in the SESAR scenario potential conflicts will be less frequent than in the current scenario although they will be more widespread over all the entire airspace. This in principle increase the complexity controllers will have to face in the SESAR scenario.

Moreover, we show that our simulator is able to tackle complexity as much as humans do in the current Air Traffic Management scenario. This will ensure that we can use the simulator to perform realistic scenario simulations to assess how the controllers will operate in the SESAR scenario.

The data we will consider below were obtained by first selecting the first day of AIRAC 334 (06 May 2010) and the LIRR ACC which covers most of Central Italy. We then generated synthetic M1 trajectories by using the Flight Plan Generator with no capacity constraints of section \ref{StratLayerModule} \cite{git_repo_model_description}. Moreover, such trajectories were de-conflicted by using the module described in section \ref{PreTactModule} \cite{git_repo_model_description}. That was done in order to discard any effect due to the resolution of possible conflicts, given the fact that in the SESAR scenario it is assumed that the flight trajectories released by the Network Manager will be conflict-free. We generated N=100 realizations of the given day. These trajectories are subsequently rectified by using the module described in section \ref{RectModule} in order to generate sets of flight-plans corresponding to different level of efficiency ranging from a low value of $E=0.973$ corresponding to the current scenario to the highest value of $E=0.999$ corresponding to the SESAR scenario. Trajectories were generated for different number of aircraft present in the airspace. These values ranged from $N_f=1500$ to $N_f=2200$. From real data, we can observe that the number of aircraft actually present in the considered airspace, given the applied filters, is approximately $N_f=1800$.



\subsection{Going from the Current Scenario to SESAR Scenario} \label{eff}

\begin{figure}  [H]
 \centering
                      \includegraphics[width=9cm]{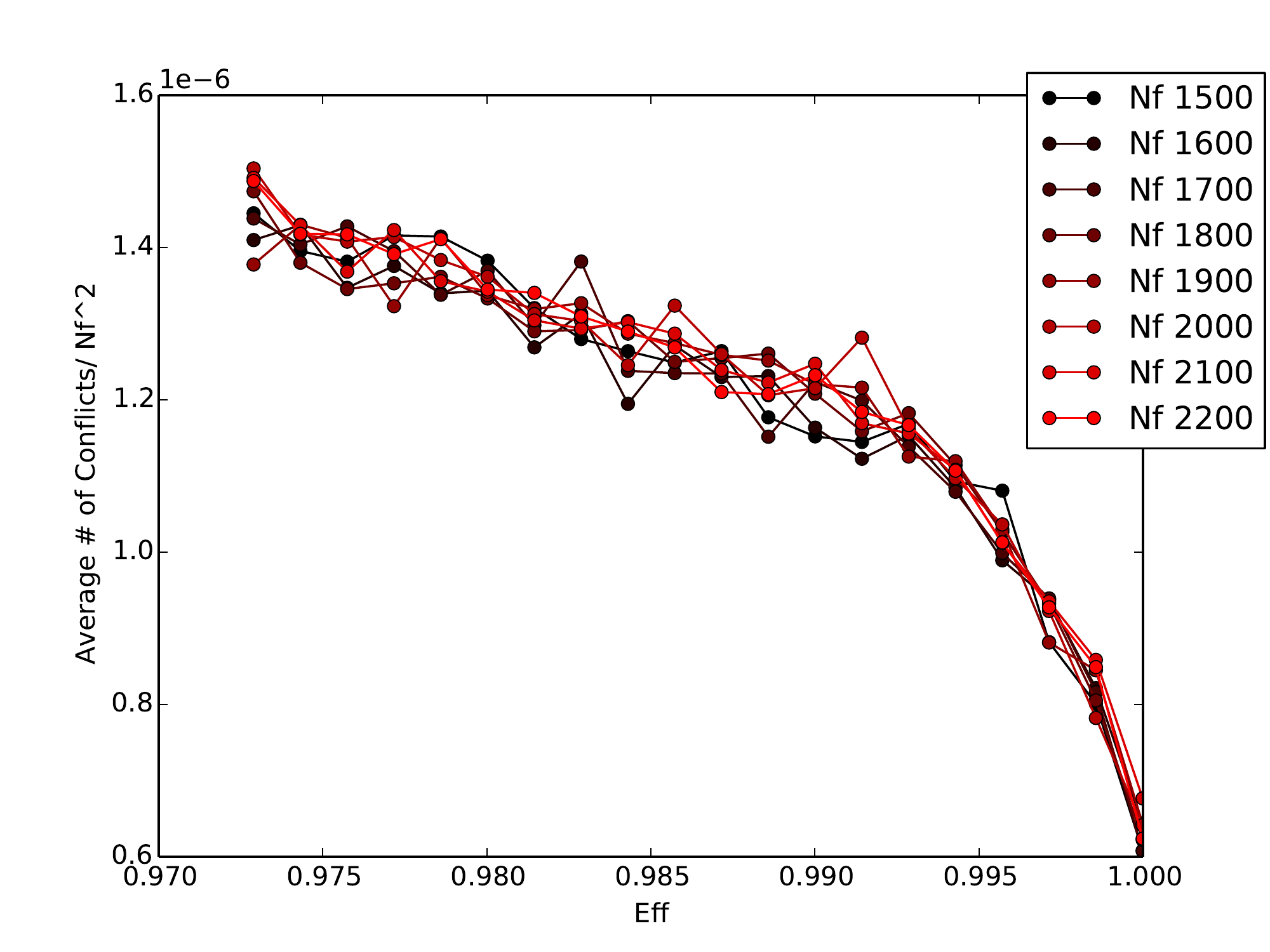} 
                      \caption{Average number of conflicts detected in the surrogate M3 flight trajectories of the LIRR ACC, for different values of efficiency (horizontal axis) and for different numbers aircraft present in the ACC (different lines in the plot). Each of the curve has been normalized with $N_f^2$ that represents the maximum possible number of conflicts in an environment with $N_f$ aircraft.}\label{res1-M3}
\end{figure}

In Fig. \ref{res1-M3} we show the average number of conflicts detected in the LIRR ACC, for different values of efficiency (horizontal axis) and for different values of the number of aircraft present in the ACC (different lines in the plot). Each of the curves has been normalized with $N_f^2$, i.e. with the expected possible number of conflicts in an environment with $N_f$ aircraft. The average number of conflicts is here measured as the average number of actions that the controller has to perform in order to avoid the conflicts detected by the Collision Module described in section \ref{TactLayerModule}. Therefore, these measures are performed on the surrogate M3 flight trajectories generated by our model. Indeed, the figure shows two interesting features: on one side we have that all curves seem to collapse in a single curve when the number of conflicts is rescaled with $N_f^2$. Moreover, the number of detected conflicts diminishes as long as efficiency increases, thus indicating that in the SESAR scenario we should observe less conflicts and therefore a smaller workload for controllers.

\begin{figure}  [H]
 \centering
                      \includegraphics[width=9cm]{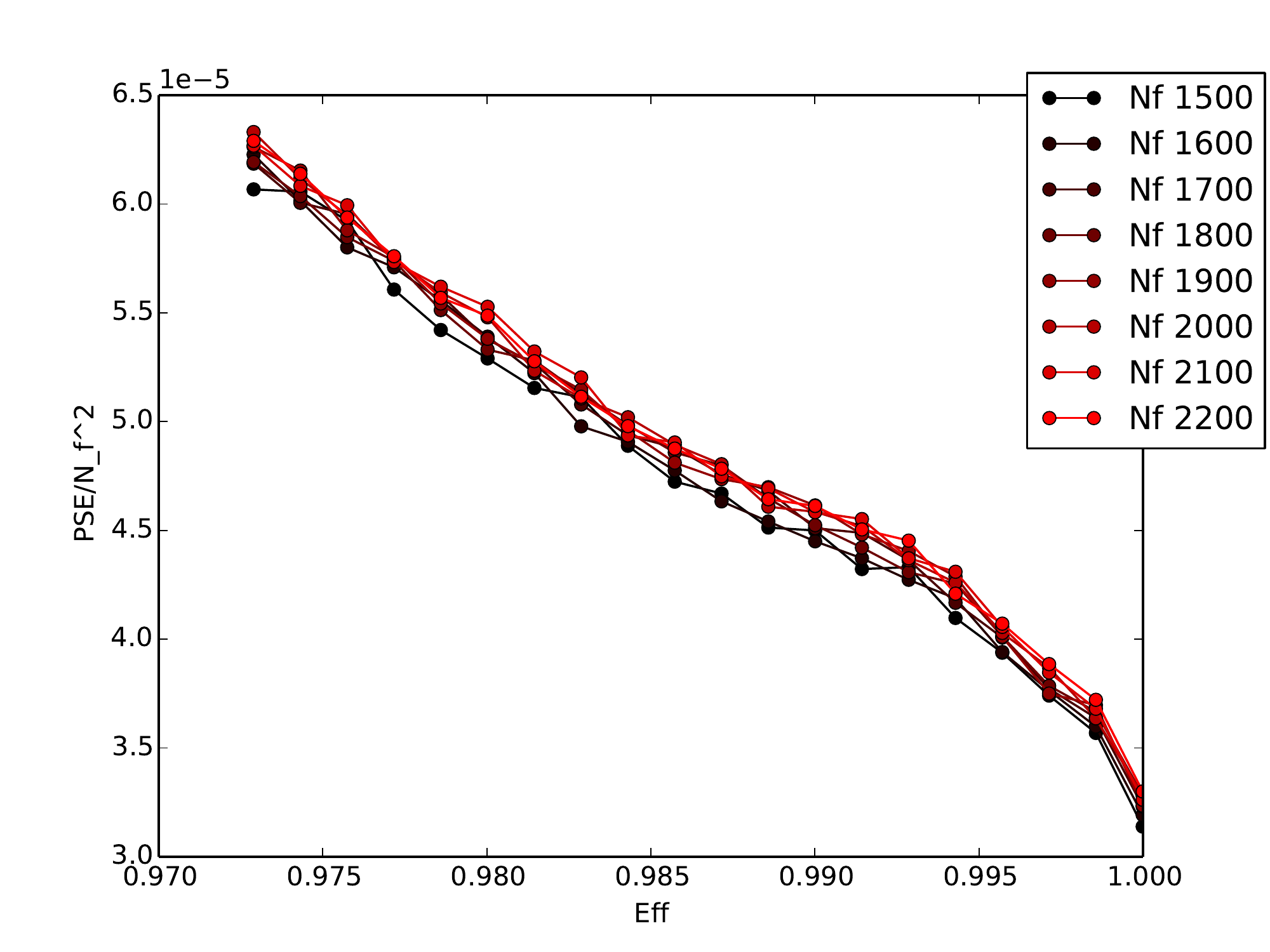} 
                      \caption{Average number of PSEs detected in the M1 flight trajectories of the LIRR ACC, for different values of efficiency (horizontal axis) and for different numbers of aircraft present in the ACC (different curves in the plot). Each of the curve has been normalized with $N_f^2$ that represents the maximum possible number of conflicts in an environment with $N_f$ aircraft.}\label{res1-M1}
\end{figure}

We have also devised a simple procedure to compute what is the expected number of possible safety events (PSE), i.e. potential conflicts, we should expect if all flights were to stick exactly to their flight plans. Indeed, since we add some noise on the departure times of the flights, the expected conflicts might not occur, and others non-expected events can take place. Moreover, we can also understand whether the fact that we observe less conflicts is already present at the level of planning or this is mainly due to the management of trajectories done by controllers. We start from the M1 de-conflicted trajectories and implement the following procedure:
\begin{itemize}
\item we perform a very fine spatial sampling of all flight trajectories. Sample points are one meter apart the one from each other.
\item starting from the original flight plans we associate to each of these sampled points a timestamp. This is done by assuming that between two navigation points the velocity of the aircraft is constant.
\item we select those sampled points $P_i^{(a)}$ in the $k$-th flight trajectory and $P_j^{(b)}$  in the $b$-th flight trajectory such that the Euclidean distance $d(P_i^{(a)},P_j^{(b)})$ between the two points is smaller than the safety threshold distance $d_{thresh}= 5$  NM.
\item we further select the points such that the times $t_i^{(a)}$ at which the $a$-th aircraft crosses $P_i^{(a)}$ and $t_j^{(b)}$ at which the $b$-th aircraft crosses $P_j^{(b)}$ are below a certain time threshold $T_{thresh}$.
\end{itemize}
By using such procedure we are able to show what are the points of the ACC that are likely to attract the controller attention as a source of possible conflicts. Of course, the PSEs thus defined are strictly depending on the $T_{thresh}$ considered. In Fig. \ref{res1-M1} we show the PSEs detected in the LIRR ACC, for different values of efficiency (horizontal axis) and for different values of the number of aircraft present in the ACC (different curves in the plot). as above, each of the shown curves has been normalized with $N_f^2$. In the figure we show the results for $T_{thresh}=5.0$ min
Also in this case, the figure shows two interesting features: on one side we have that all curves seem to collapse in a single curve when the number of PSEs is rescaled with $N_f^2$ and the number of PSEs diminishes as long as efficiency increases, thus indicating that in the SESAR scenario we should expect less conflicts and therefore a smaller workload for controllers.

\begin{figure}  [H]
 \centering
                      \includegraphics[width=9cm]{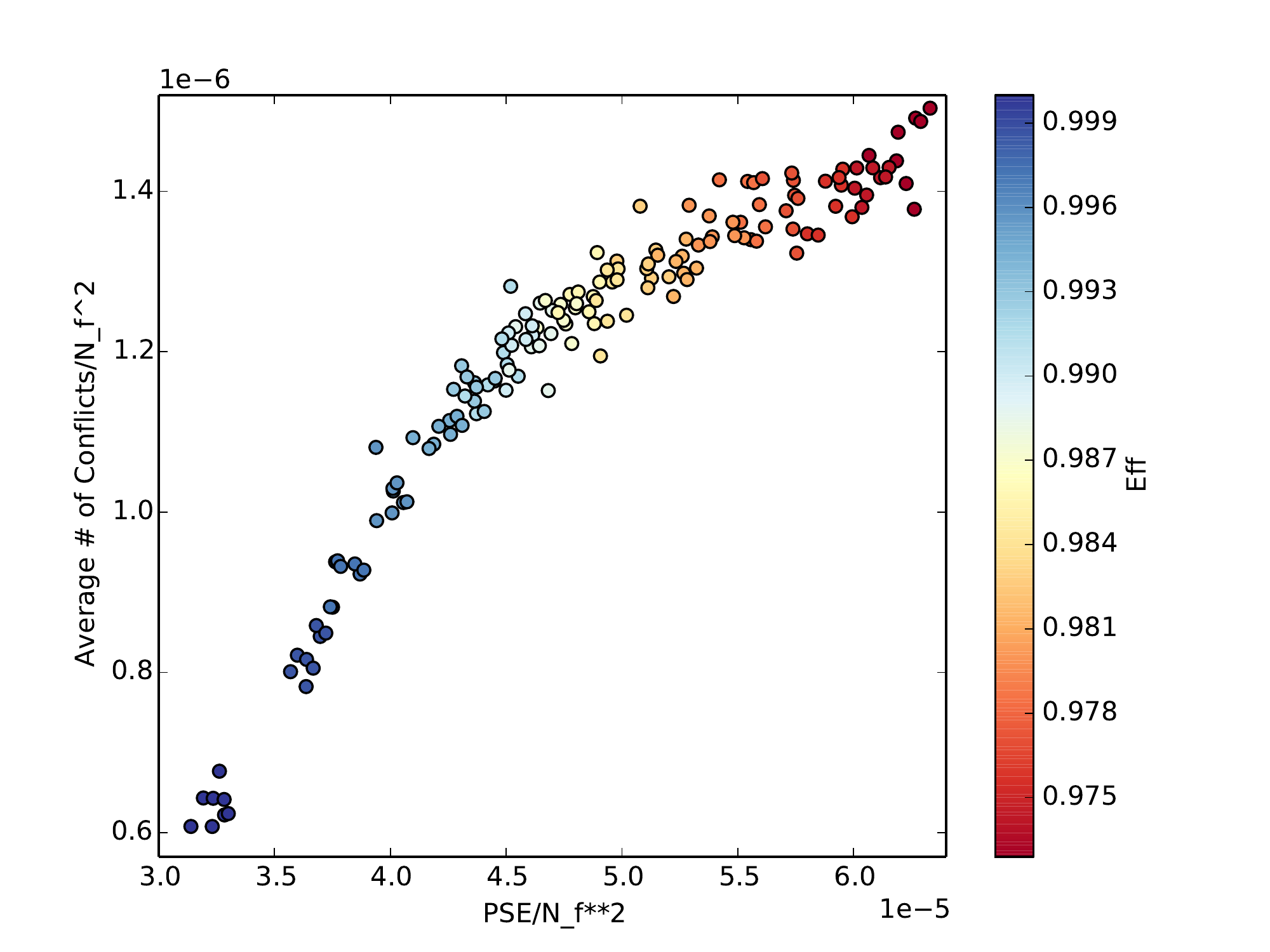} 
                      \caption{Scatter plot of the average number of conflicts detected in the surrogate M3 flight trajectories versus the average number of possible safety events (PSE) of the LIRR ACC. Different points represent different values of efficiency and different values of the number of aircraft present in the ACC.}\label{res1-scattplot}
\end{figure}

In Fig. \ref{res1-scattplot} we show a scatter-plot between the normalized PSEs detected from the M1 files with $T_{thresh}=5.0$ min (horizontal axis) and the normalized number of conflicts detected from the surrogate M3 files (vertical axis) for different values of efficiency. The figure shows the existence of two different regimes. For values of efficiency close to unity the curve can be fitted with a linear relationship whose slope is of the order of $0.05$, while for lower values of efficiency we have a linear relationship whose slope is of the order of $0.01$. In any case, the fact that the curve is steeper for high values of efficiency indicates that a small variation in the PSEs translates into a larger variation of the number of detected conflicts, thus indicating that the SESAR scenario might reveal to be less flexible to accommodate variations in the planning of the trajectories.

\subsection{Heterogeneity} \label{hetero}

The above results show that in the SESAR scenario one should expect to observe less conflicts than in the current scenario. We want now to investigate what are their spatial locations. In fact, the main reason for having a navigation point grid is that this helps the controllers in monitoring the air traffic only in specific portions of the airspace. We are therefore interested in understanding whether or not this feature will be maintained in the SESAR scenario.

\begin{figure}  [H]
 \centering
                      \includegraphics[width=5.8cm]{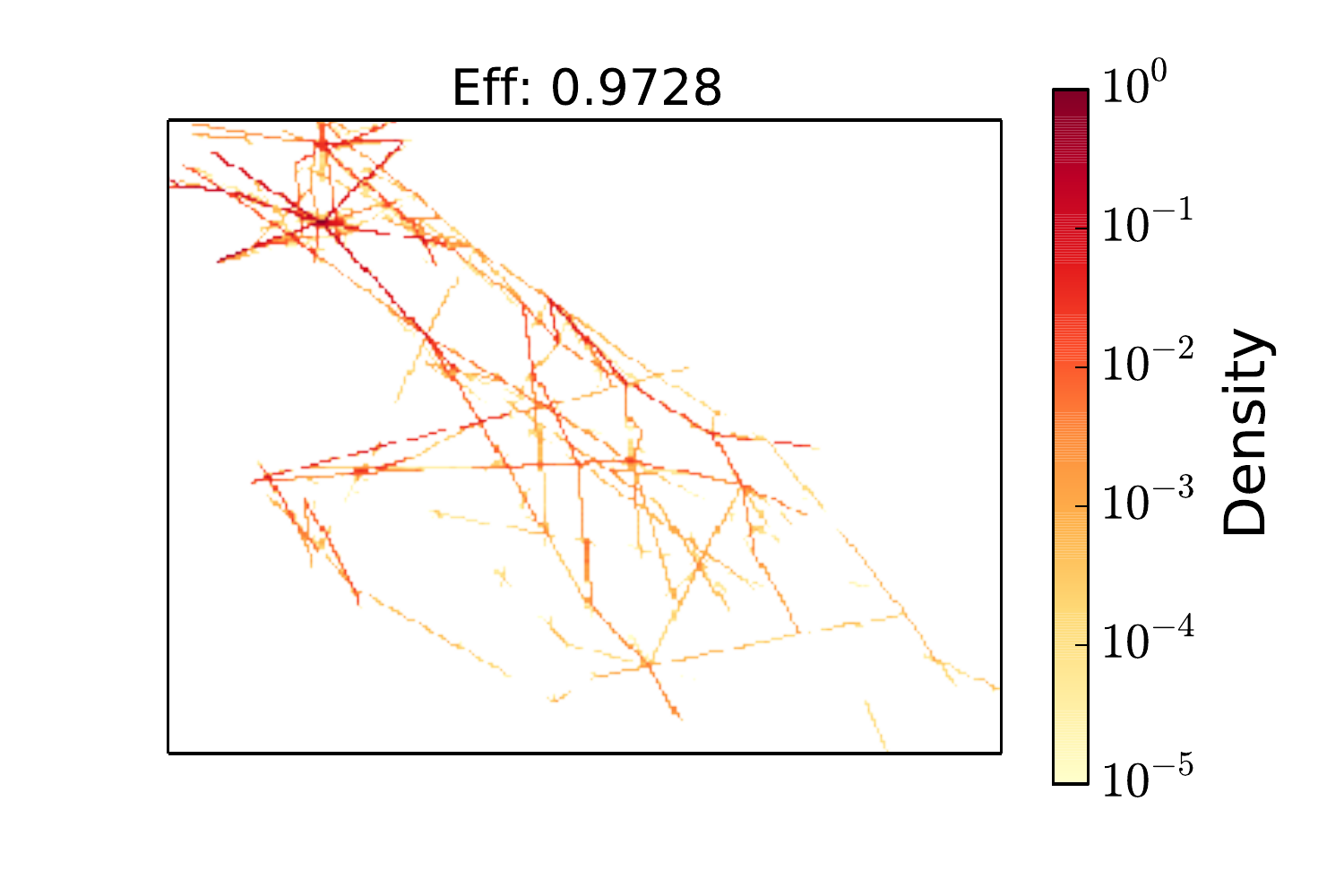} 
                      \includegraphics[width=5.8cm]{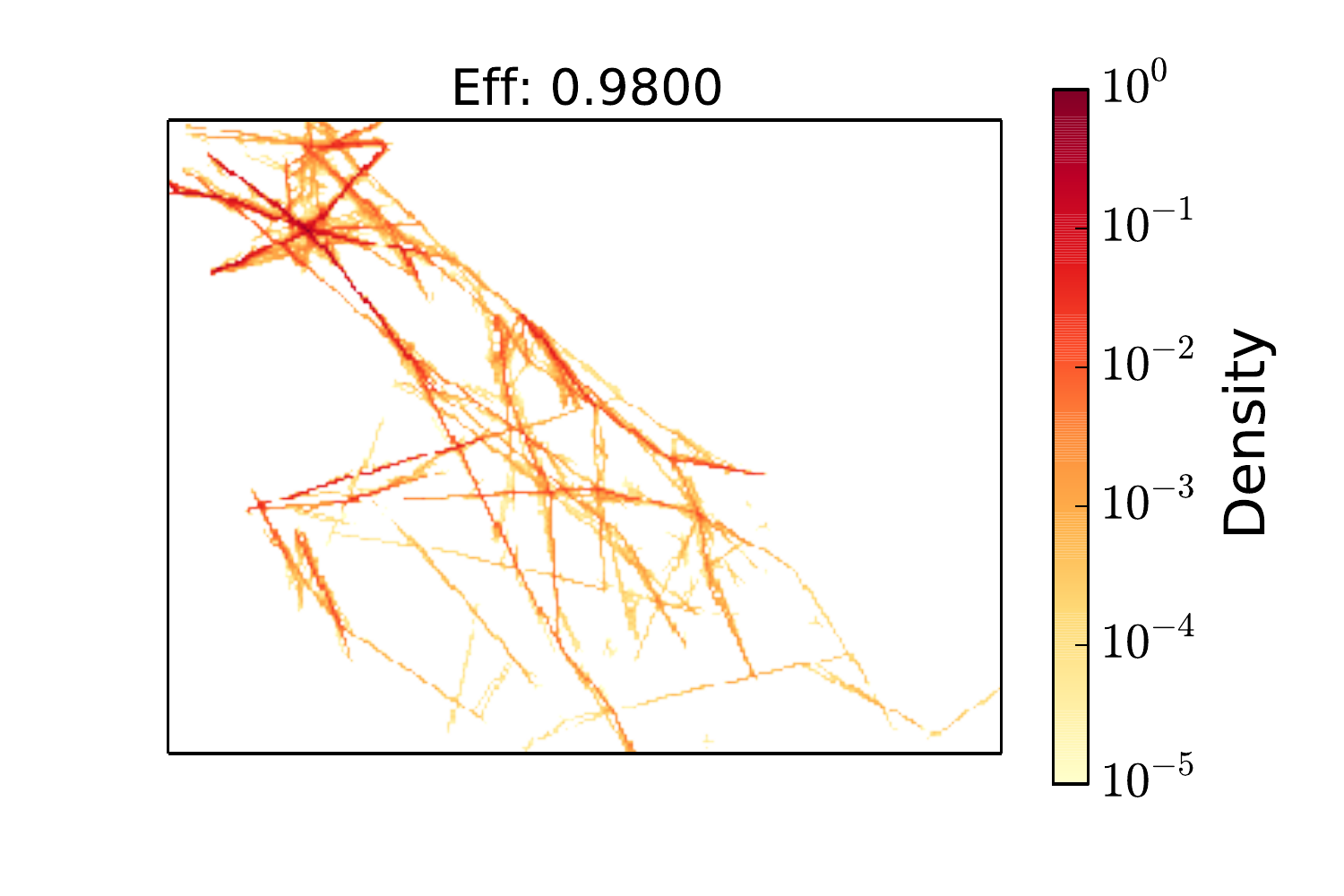} 
                      \includegraphics[width=5.8cm]{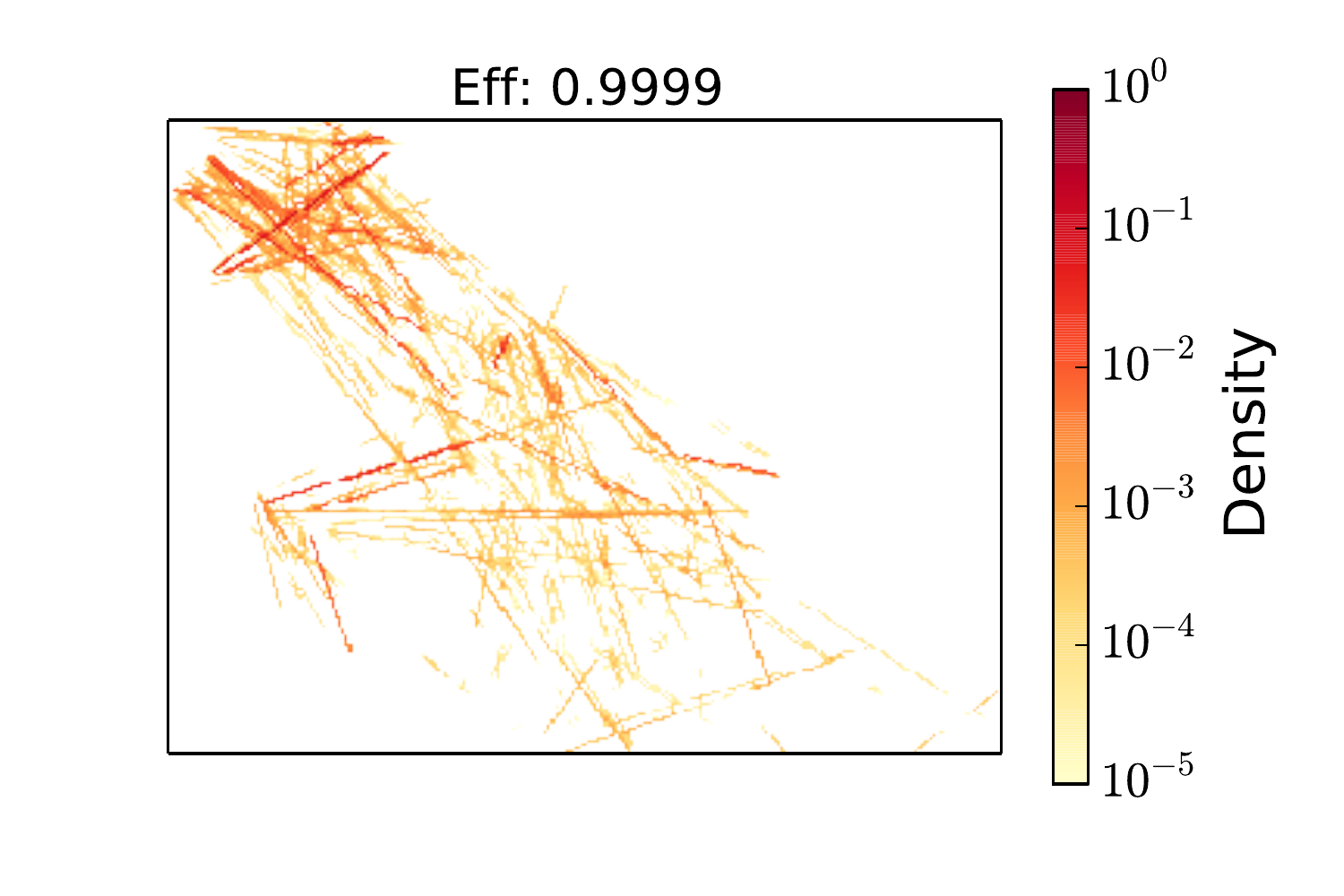} 
                      \caption{Density map of the PSEs detected when considering three different values of efficiency and $T_{thresh}=5.0$ min in the LIRR ACC. In the left panel we show the PSEs detected starting from the real M1 trajectories, i.e. of $E=0.973$. In the right panel we show the PSEs detected starting from the M1 trajectories corresponding to the SESAR scenario, i.e. with an efficiency value of $E=0.999$. In the central panel we show the PSEs detected starting from the M1 trajectories corresponding to the intermediate value of efficiency $E=0.980$. To enhance readability, we first take the logarithm of the number of PSEs and then we normalize by dividing all values of the maximum value found in the three graphs (which is reached in the left panel).}\label{res1-PSE}
\end{figure}

In Fig. \ref{res1-PSE} we show a density map of the PSEs detected when considering three different values of efficiency and $T_{thresh}=5.0$ min. In the left panel we show the PSEs detected starting from the real M1 trajectories, which corresponds to an efficiency value of $E=0.973$. In the right panel we show the PSEs detected starting from the M1 trajectories corresponding to the SESAR scenario, i.e. with an efficiency value of $E=0.999$. In the central panel we show the PSEs detected starting from the M1 trajectories corresponding to the intermediate value of efficiency $E=0.980$. As expected, as long as efficiency increases the possible conflicts are more spread all over the ACC, rather than being concentrated in specific regions. In fact, flight trajectories are more distributed over the entire airspace and therefore the probability of conflicting is smaller. This explains why the number of detected conflicts diminishes as long as efficiency increases. This also implies that the controller activity in the SESAR scenario will change, moving from a situation where he/she has to give attention to a high number of conflicts concentrated in specific points to a situation where he/she will have to manage less conflicts spread over a much larger portion of the airspace.

Indeed, controllers are obviously sensitive not only to the number of conflicts or the number of flights, but to other factors. In fact, to our knowledge there is no consensus about what is a complex situation for controllers. In literature \cite{Histon2002}  there are examples of several metrics that capture specific aspects of the complexity typically faced by the controllers. These metrics are quite diverse and are linked to time to conflicts, distances between aircraft, geometry of conflicts, etc. They are computable with the planned trajectories, which allows to see at each point in time what is the expected complexity for the controller. On the other hand, we also have our own measure of complexity, directly coming from the model. In fact, the number of actions done by the controllers can be viewed as a complexity measure where our (super-)controller gradually meets more complex situations and thus makes more actions to solve the conflicts. As a consequence, we are here interested in investigating whether the number of actions can be related to the complexity metrics already known in literature. 

In order to do so, we computed the metrics of \cite{Laudeman1998,Sridhar1998} and \cite{Chatterji2001} by using our planned trajectories. Specifically, given the generated trajectories we also computed the number of actions necessary for solving conflicts. Table \ref{tab:complexity} shows the average values over one single day of these metrics. Data shown are also averaged over the $N \sim 100$ considered realizations. The meaning of the metrics can be found in Appendix \ref{app1}. Quite strikingly, despite the results of Fig. \ref{res1-PSE}, most of the metrics decrease in the new scenario. The metrics which increase are all linked to the altitude, essentially because with free routing as we designed it the flights are always descending or ascending, but never stay at a constant altitude. 
Consistently with most of the metrics, the number of actions of the controller also drops, as also indicated in Fig. \ref{res1-M1} and Fig. \ref{res1-M3}. Note also that the metric C16, which captures the geometry of potential conflicts, also drops a lot (by 64\%). This counter-intuitive results is probably due to the fact that this metric scales with the number of conflicts. Since the latter drops a lot in the new scenario, the possible higher complexity due to geometry disappears with the number of conflicts. Another explanation might be that since conflicts are more spread all over the considered airspace, see Fig. \ref{res1-PSE}, on average aircraft stay at a larger distance, which makes the topological constraints less stringent.
\begin{table}[H]
\begin{center}
\begin{tabular}{l||c|c|c|c}
\hline
{} &    Low Eff &   High Eff &  Norm. diff. &     T test \\
\hline
\hline
na\_v   &   0.051049 &   0.021128 &    -0.414553 &      Diff. \\
na\_h   &   0.048543 &   0.022056 &    -0.375179 &      Diff. \\
C1     &   1.747375 &   1.702682 &    -0.012954 &      Diff. \\
C2     &   0.267801 &   0.288694 &     \textbf{0.037543} &      Diff. \\
C3     &   0.290026 &   0.307920 &     \textbf{0.029924} &      Diff. \\
C4     &   0.337407 &   0.293016 &    -0.070414 &      Diff. \\
C5     &   0.003366 &   0.003196 &    -0.025883 &  Not Diff. \\
C6     &   0.000194 &   0.000191 &    -0.007885 &  Not Diff. \\
C7     &   0.001828 &   0.001920 &     \textbf{0.024651} &      Diff. \\
C8     &   0.000107 &   0.000064 &    -0.252736 &      Diff. \\
C9     &   0.003678 &   0.016946 &     \textbf{0.643301} &      Diff. \\
C10    &   0.000122 &   0.000070 &    -0.268582 &      Diff. \\
C11    &   0.039245 &   0.032656 &    -0.091634 &      Diff. \\
C12    &   0.000002 &   0.000002 &    -0.012337 &  Not Diff. \\
C13    &   0.041139 &   0.019182 &    -0.363997 &      Diff. \\
C16    &   0.000472 &   0.000101 &    -0.646889 &      Diff. \\
N      &  34.967815 &  34.053641 &    -0.013245 &      Diff. \\
HC     &   3.244316 &   0.000000 &    -1.000000 &      Diff. \\
AC     &  15.659597 &  17.028818 &     \textbf{0.041887} &      Diff. \\
MD5    &   0.807553 &   0.467555 &    -0.266642 &      Diff. \\
MD10   &   0.673807 &   0.446676 &    -0.202708 &      Diff. \\
CP25   &   1.683513 &   1.412701 &    -0.087465 &      Diff. \\
CP40   &   2.080441 &   1.949608 &    -0.032464 &      Diff. \\
CP75   &   7.737588 &   6.996779 &    -0.050278 &      Diff. \\
na\_tot &   0.099592 &   0.043184 &    -0.395084 &      Diff. \\
\hline
\end{tabular}
\end{center}
\caption{Values of different complexity metrics in the low efficiency scenario and the high efficiency scenario. The third column indicates the normalized difference between both, and the last one the result of a T-test to see if the metrics are statistically different in both scenarios. The values presented are computed over one day. The first two rows and the last ones are coming from simulations, whereas the other ones are computed over planned trajectories only. More specifically, na\_h and na\_v are the number of actions that the controller does on the horizontal and vertical axes, whereas na\_tot is just the total. The other metrics are coming from papers of the literature on the subject and their meaning can be found in appendix \ref{app1}. We have highlighted in bold the metrics which increase in the new scenario.} \label{tab:complexity}
\end{table}

In summary, it seems that the complexity indicators are decreasing overall in the new scenario. As a consequence, we have a strong indication that the situation would be less complex for humans to handle.  Moreover, the indicators of complexity coming from the algorithm are going in the same direction. In order words, it seems that our virtual super-controller and some real controllers would react in the same way to changing conditions of traffic. If this point is confirmed, we could use safely use our simulator to predict the complexity of situations in new environment, including free-routing.

\subsection{Complexity Scaling} \label{scaling}

Before going further in the use of these complexity metrics, it is worth understanding better what is the relationship between the complexity and its most simple component, the density. 

To investigate this issue we perform simulations on the same airspace but with changing capacities for the sectors. We first produce a fixed number of trajectories with the strategic layer, and then we change the capacities in three different ways:
\begin{itemize}
\item In the first scenario, we decrease uniformly the capacities of all sectors.
\item In the second one, we ``impair'' severely three central sectors, increasing the capacities of the surrounding sectors to have the same average capacity. Then we decrease all capacities uniformly like in the previous point.
\item The last one is the witness in which we remove the capacity constraint and change the number of flights submitting a flight plan.
\end{itemize}
After that, we use the tactical layer to solve all conflicts in each simulations and we track the number of actions of the controllers. Figure \ref{fig:pca} shows the output. For each of the scenarios, we performed a power-law regression with the function $N_f \mapsto bN_f^a$. For the red line -- without capacity constraints -- we obtain $a = 2.0 \pm 9.0\,10^{-5}$ and $b = 8.4\, 10^{-5} \pm 3.0\,10^{-11}$, i.e. a pure quadratic law. This is expected, because the number of conflicts should scale with the number of pairs of aircraft, i.e. $\sim N_f(N_f-1)/2 \sim N_f^2$.
\begin{figure}[htbp]
\begin{center}
	\includegraphics[width=0.65\textwidth]{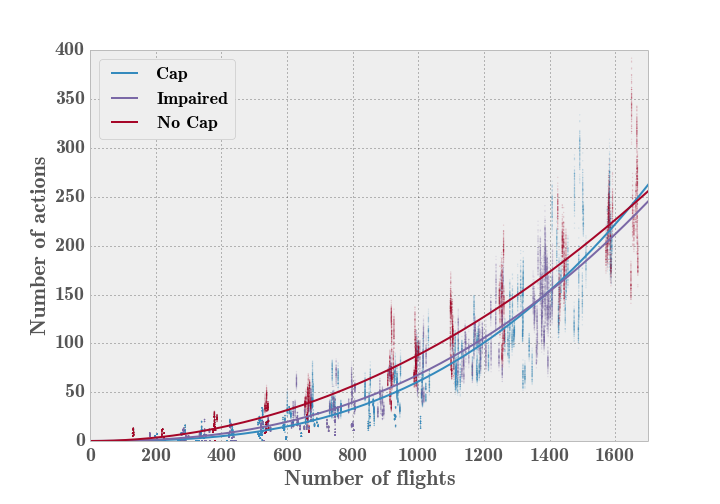}
\end{center}
\caption{Different scaling for different scenarios. The scatter plots have been obtained with a uniform reduction of the capacities (light blue), with some sectors severely impaired (violet) and without any capacity (red). The solid lines are the result of power law regressions for each set of data (see text).} \label{fig:pca}
\end{figure}

Interestingly, the regression for the two other scenarios yield scalings. With the same regression, the violet line yields: $a= 2.4 \pm 1. 10^{-4}$ and $b=3.6 10^{-6}\pm 8. 10^{-14}$ and the blue line one gives: $a= 2.75 \pm 2. 10^{-4}$ and $b=3.4 10^{-7}\pm 1. 10^{-15}$. In other words, they are clearly displaying super-quadratic behaviors. But before explaining why, we comment on the fact that despite this behavior, these two cases usually need less actions than the capacity-free case for the same number of flights. This is due to the fact that capacities tend to spread the flights during the day. Hence the concentration in time of flights decreases during peaks, which decreases the number of potential conflicts (flights are flying at different times).

The same argument explains the super-quadratic behavior. Indeed, due to our experimental procedure, when the number of flights increases, it means that the capacities are less binding, since we keep the number of flights fixed as input to the strategic layer. Hence, when the number of flights increases, the number of potential conflicts increases more quickly than  $N_f^2$, because more flights are flying at similar times. Finally, note that the same kind of simulations in free-routing yields the same kind of results (not shown here). 

The conclusion is that sectors play a major role the complexity of the airspace, and that the complexity heavily depends on the pairs of flights simultaneously present in the airspace. All the variance is not explained by the density though, and other factors are important, especially to human controllers.

\subsection{Complexity Metrics} \label{compmet}

In this section we will assess the capability of our simulator to tackle complexity as much as humans do in the current Air Traffic Management scenario. This will be done by considering the complexity indicators already introduced above and comparing their values obtained in the case of our simulator and in the case of human decisions. Having clarified this aspect, in the next section we will therefore use the simulator to assess how the controllers will operate in the SESAR scenario.

\subsubsection{Structure of the correlation matrix}

In order to address such question, we first consider the correlation matrix computed starting from the above metrics. Specifically, we compute the Pearson correlation coefficients over one day of data with a resolution of two minutes. The correlations are presented in tables \ref{tab:correlations_first_batch} and \ref{tab:correlations_second_batch}. 

\begin{table}[htbp]
\begin{center}
\begin{tabular}{l||r|r|r|r|r|r|r|r|r|r|r|r|r|r}
{} &    C1 &    C2 &    C3 &    C4 &    C5 &    C6 &    C7 &    C8 &    C9 &   C10 &   C11 &   C12 &   C13 &   C16 \\
\hline
\hline
C1  &  1.00 &  0.63 &  0.63 &  0.63 &  0.91 &  0.44 &  0.94 &  0.82 &  0.65 &  0.82 &  0.99 & -0.12 &  0.69 &  0.92 \\
C2  &  0.63 &  1.00 &  0.77 &  0.72 &  0.80 &  0.76 &  0.74 &  0.50 &  0.47 &  0.48 &  0.62 & -0.06 &  0.39 &  0.60 \\
C3  &  0.63 &  0.77 &  1.00 &  0.90 &  0.82 &  0.82 &  0.77 &  0.48 &  0.48 &  0.47 &  0.62 & -0.03 &  0.40 &  0.60 \\
C4  &  0.63 &  0.72 &  0.90 &  1.00 &  0.82 &  0.81 &  0.77 &  0.48 &  0.49 &  0.46 &  0.62 & -0.02 &  0.40 &  0.60 \\
C5  &  0.91 &  0.80 &  0.82 &  0.82 &  1.00 &  0.66 &  0.98 &  0.72 &  0.65 &  0.71 &  0.91 & -0.09 &  0.60 &  0.85 \\
C6  &  0.44 &  0.76 &  0.82 &  0.81 &  0.66 &  1.00 &  0.59 &  0.33 &  0.35 &  0.32 &  0.43 & -0.00 &  0.27 &  0.42 \\
C7  &  0.94 &  0.74 &  0.77 &  0.77 &  0.98 &  0.59 &  1.00 &  0.75 &  0.66 &  0.74 &  0.93 & -0.12 &  0.62 &  0.88 \\
C8  &  0.82 &  0.50 &  0.48 &  0.48 &  0.72 &  0.33 &  0.75 &  1.00 &  0.58 &  0.95 &  0.81 & -0.10 &  0.56 &  0.75 \\
C9  &  0.65 &  0.47 &  0.48 &  0.49 &  0.65 &  0.35 &  0.66 &  0.58 &  1.00 &  0.60 &  0.64 & -0.08 &  0.43 &  0.57 \\
C10 &  0.82 &  0.48 &  0.47 &  0.46 &  0.71 &  0.32 &  0.74 &  0.95 &  0.60 &  1.00 &  0.81 & -0.09 &  0.59 &  0.74 \\
C11 &  0.99 &  0.62 &  0.62 &  0.62 &  0.91 &  0.43 &  0.93 &  0.81 &  0.64 &  0.81 &  1.00 & -0.09 &  0.69 &  0.91 \\
C12 & -0.12 & -0.06 & -0.03 & -0.02 & -0.09 & -0.00 & -0.12 & -0.10 & -0.08 & -0.09 & -0.09 &  1.00 & -0.08 & -0.12 \\
C13 &  0.69 &  0.39 &  0.40 &  0.40 &  0.60 &  0.27 &  0.62 &  0.56 &  0.43 &  0.59 &  0.69 & -0.08 &  1.00 &  0.63 \\
C16 &  0.92 &  0.60 &  0.60 &  0.60 &  0.85 &  0.42 &  0.88 &  0.75 &  0.57 &  0.74 &  0.91 & -0.12 &  0.63 &  1.00 \\
\end{tabular}
\caption{Pearson correlation coefficients for the complexity metrics found in \cite{Chatterji2001} (first batch).} \label{tab:correlations_first_batch}
\end{center}
\end{table}

\begin{table}[htbp]
\begin{center}
\begin{tabular}{l||r|r|r|r|r|r|r|r}
{} &    N &   HC &   AC &  MD5 &  MD10 &  CP25 &  CP40 &  CP75 \\
\hline
\hline
N    & 1.00 & 0.99 & 1.00 & 0.95 &  0.93 &  0.96 &  0.96 &  0.97 \\
HC   & 0.99 & 1.00 & 0.99 & 0.94 &  0.93 &  0.95 &  0.96 &  0.97 \\
AC   & 1.00 & 0.99 & 1.00 & 0.95 &  0.94 &  0.96 &  0.96 &  0.97 \\
MD5  & 0.95 & 0.94 & 0.95 & 1.00 &  0.96 &  0.98 &  0.98 &  0.98 \\
MD10 & 0.93 & 0.93 & 0.94 & 0.96 &  1.00 &  0.97 &  0.97 &  0.97 \\
CP25 & 0.96 & 0.95 & 0.96 & 0.98 &  0.97 &  1.00 &  0.98 &  0.99 \\
CP40 & 0.96 & 0.96 & 0.96 & 0.98 &  0.97 &  0.98 &  1.00 &  0.99 \\
CP75 & 0.97 & 0.97 & 0.97 & 0.98 &  0.97 &  0.99 &  0.99 &  1.00 \\
\end{tabular}
\caption{Pearson correlation coefficients for the complexity metrics found in \cite{Laudeman1998} (second batch).} \label{tab:correlations_second_batch}
\end{center}
\end{table}

We first notice that these metrics are very far from being independent from each other. Except for the case of C12, all metrics shows quite high correlations between them. In particular, the metrics N, HC, etc. are very correlated with each other. In fact, this is expected, since these metrics are not normalized with the square of the traffic, i.e. they all scale at least as $\sim N_f^2$, as shown in section \ref{scaling}.  Hence, their correlations just reflect the daily pattern of traffic. For the same reason, the choice of the density as the first relevant metric in each batch (C1 and N) is not necessarily the most relevant. Note however that the correlation between the metrics can also have other sources, but normalizing the metrics with adequate scaling allows precisely to better see these secondary effects. In any case, this high degree of correlation poses the problem of the real causes of the complexity for controllers. At best, the use of these metrics can be viewed as an inadequate ``base'' for the description of the complexity metrics space. 

\subsubsection{Reducing the dimensionality of the complexity}

Statistically speaking, the variance of all these variables can be explained with a smaller number of ``hidden'' variables, which can be interpreted as the real descriptors of the complexity. With this idea in mind, we performed a Principal Component Analysis (PCA) \cite{Pearson1901}, aiming at discovering these hidden variables and reduce the dimensionality of the problem. For this analysis, we used the first batch of metrics, i.e. metrics from C1 to C16. Since these metrics are already properly normalized, this will give cleaner results. Moreover, to some extent, they include the second batch of metric through quite high correlations between them (not shown here). 

After performing the PCA on the correlation matrix of Eq. \ref{tab:correlations_first_batch}, we chose to keep the first four components, which account for about 88\% of the variance. Their compositions in terms of the initial variables C1 -- C16 are shown in Fig. \ref{fig:pca}. As usual with PCA, the difficult part is to give a physical interpretation to the new variables. This can be very instructional and some suggestions are made thereafter.

The first component is clearly related to a scaling not adequately captured by the normalizations of the metrics, i.e. a global dependence on the traffic. The second component might be related to a differential complexity between  horizontal and vertical components, since most of the negative weights are related to vertical metrics whereas the positive weights are associated to more horizontal metrics. The third metric is almost purely the metric C12, which was expected from the correlation matrix in table \ref{tab:correlations_first_batch}, since it has very low correlations with other metrics. Finally, the last component is clearly related to the speed of aircraft: C9, related to the inverse of the minimum distance between aircraft, has a negative weight, whereas C13, related to the minimum time to conflict. Indeed, greater speed and fix time horizon for the controller implies smaller C9 and bigger C13. 

\begin{figure}[H]
\begin{center}
                       \includegraphics[width=0.45\textwidth]{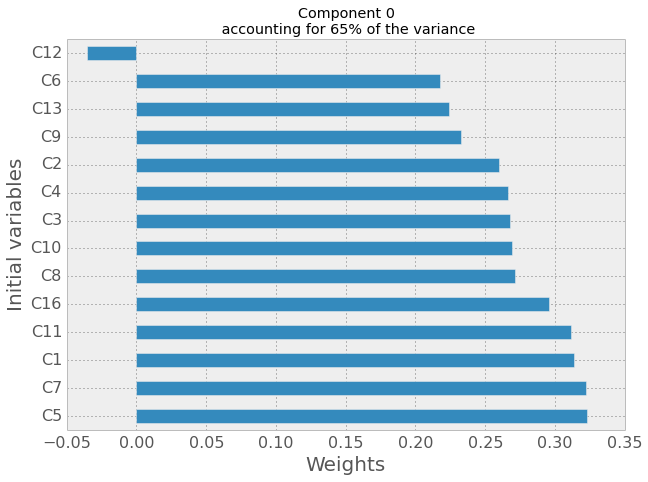}
                       \includegraphics[width=0.45\textwidth]{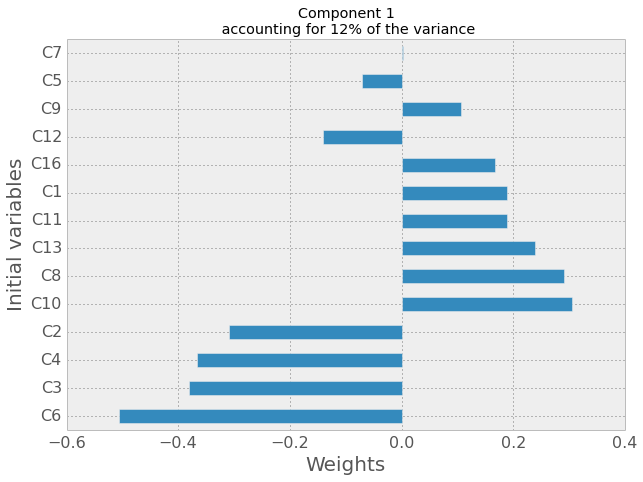}
                       \includegraphics[width=0.45\textwidth]{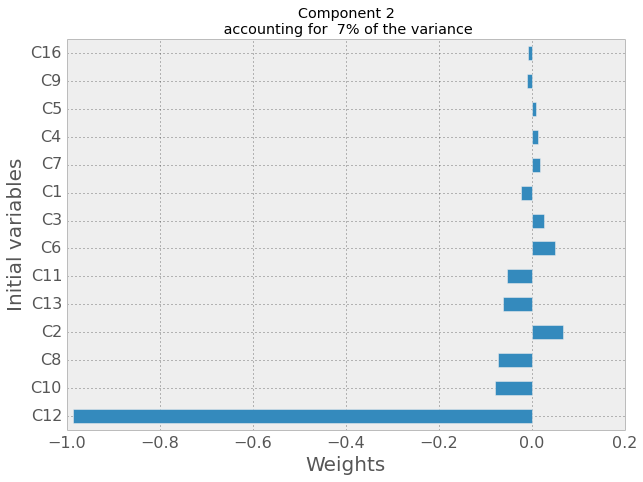}
                       \includegraphics[width=0.45\textwidth]{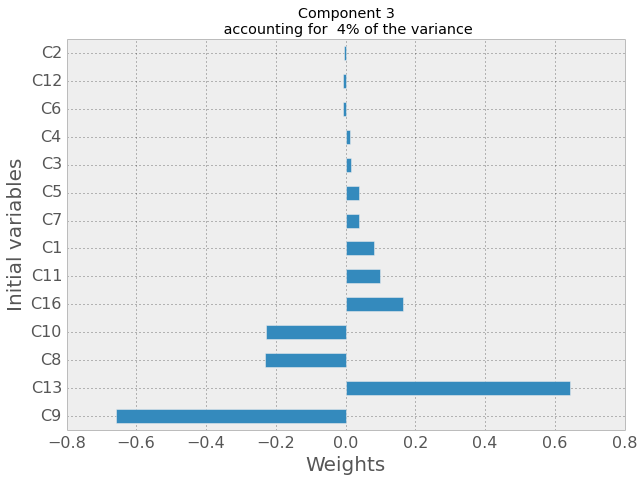}
\end{center}
\caption{Composition of the first four eigenvectors of the PCA in the current scenario. } \label{fig:pca}
\end{figure}

Note that all these results have been obtained with non-rectificated trajectories, i.e. in the current scenario. However, one might go further and check whether or not the determinants of complexity will be the same in the current and the SESAR scenario. Specifically, one might ask whether the four eigenvectors of Fig. \ref{fig:pca} will be the same also in the SESAR scenario. A full discussion of this issue goes beyond the scope of this paper, but we included in \ref{annex:pca_SESAR} the results of the same procedure than above on straight trajectories, i.e. in the new scenario. Preliminary results show that the components are not very different with respect to the current scenario. On one hand, this is reassuring, because it means that the metrics have the same kind of cross-dependence and thus that we can safely extrapolate the current complexity measures to the new scenario. On the other hand, the fact that they are slightly different is expected and this is exactly why our model could be beneficial to forecast the real complexity of the situation. In fact, the model is based on some kind of cognitive modelling rather than relying on assumptions on complexity. 

\subsubsection{Explaining the actions of the simulator with the new variables} \label{pippo}

Our next step is to discover how well these new combined metrics can explain the variation in the number of actions of our controllers. As stated previously, the idea is that the number of actions is a measure of the complexity for our algorithm. 

In order to do this, we perform and Ordinary Least Square regression on the data \cite{J.N.Miller2005}. The explicatory variables are the four components isolated above and the variable to explain is the number of action (na\_{tot}) of our controller. The result of the fit is presented in table \ref{tab:OLS}. The fitting procedure is quite successful, with a $R^2$ of 0.817. The weights for each variables are very well defined since the t-tests return p-values smaller than $0.5\%$. Note also that the first two components are much more important to explain the variations than the two others, which are not negligible nevertheless. This table shows that with only four complexity variables, we are able to forecast quite well the behavior of the synthetic (super-)controller of our simulator. 
\begin{table}[H]
\begin{center}
\begin{tabular}{c||c|c|c|c}
Comp.	& Weights	&	Std. err.	&	$P>|t|$	& 95\% CI  \\
\hline
\hline
0		& 0.2803	&	0.005		& 0.000	 	& [0.270, 0.291]\\
1		& 0.2431	& 0.012			& 0.000		& [0.219, 0.267]\\
2		& -0.0465	& 0.016			& 0.004		& [-0.078, -0.015]\\
3		& 0.0684	& 0.021			& 0.001		& [0.027, 0.110]\\
\end{tabular}
\caption{Ordinary Least Square method used to make a regression of the number of actions in our model against the four components found during the PCA. The first column present the weights, the second one the standard error, the third one a t-test for the coefficient, and the last one the 95\% confidence interval. Note that the variables have been standardized (i.e. we subtract the mean and divide by the standard deviation). The fit has a $R^2$ of 0.817.} \label{tab:OLS}
\end{center}
\end{table}

\begin{figure}[H]
\begin{center}
                         \includegraphics[width=0.65\textwidth]{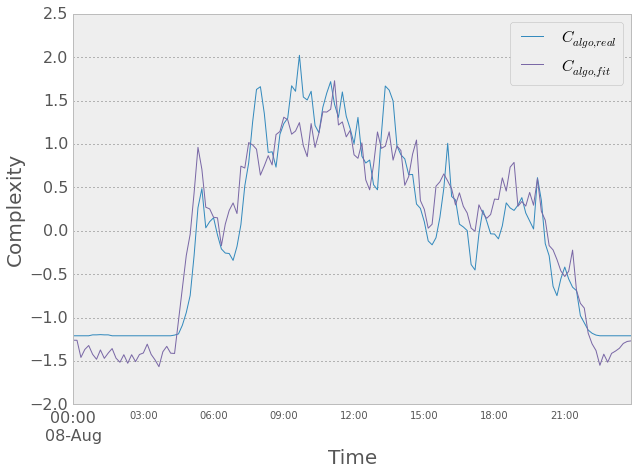}
\end{center}
\caption{Variation of four complexity metrics with time of the day. See text for more details. The data has been re-sampled every 10 minutes for visibility.} \label{fig:final_met1}
\end{figure}

Finally, in Fig. \ref{fig:final_met1} we show the variations over the day of two metrics $C_{algo, real}$ and $C_{algo, fit}$ . Specifically, $C_{algo, real}$ is the actual number of actions of our algorithm, while $C_{algo, fit}$ is the  number of actions computed by only considering our own regression of the number of actions versus the four components we have isolated with the PCA. The correlation between these two curves is 0.904, thus confirming that using the four PCA components gives a good approximation of the controllers' workload.

\subsubsection{Complexity for real controllers}

The final step for us is to compare these results with the real complexity experienced by human. For this, we base our analysis on the work of Refs. \cite{Chatterji2001,Laudeman1998}. In these two papers, the authors managed to isolate two sets of number weighting the metrics N, HC, etc. , i.e. the metrics included in the second batch. The first set is a subjective weighting scheme made by three controllers: they think for instance that the heading changes are approximatively twice as complex as the density itself.  The second set of weights comes from a regression of complexity recorded in live by controllers against the metrics, computed every 2 minutes. This gives in a way a more ``unconscious'' assessment of complexity for controllers. 

\begin{figure}[H]
\begin{center}
                      \includegraphics[width=0.65\textwidth]{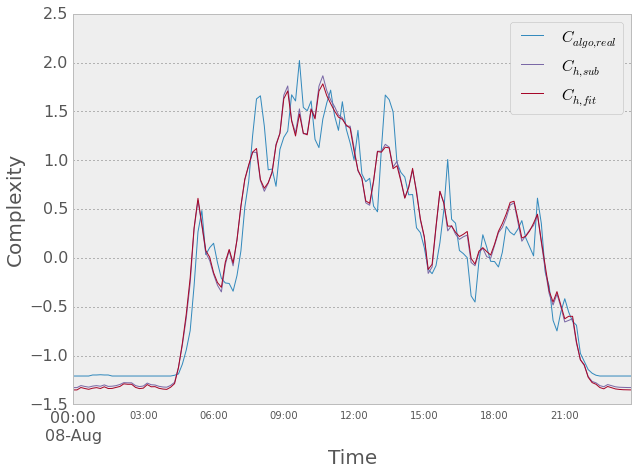}
\end{center}
\caption{Variation of four complexity metrics with time of the day. See text for more details. The data has been re-sampled every 10 minutes for visibility.} \label{fig:final_met}
\end{figure}

In our investigation, we take both these sets of weights in order to recompute an approximation of the complexity of the situations \textbf{that a human would experience should she/he control our trajectories}. In figure \ref{fig:final_met} we show the variations over the day of the $C_{algo, real}$ metrics described above together with other two metrics called $C_{h, sub}$ and $C_{h, fit}$. Specifically,  $C_{h, sub}$   is the number of actions done by the controller and computed by using the subjective weights over the set of metrics N, HC, etc. while $C_{h, fit}$ is the number of actions done by the controller and computed by using the ``regression'' weights over the set of metrics N, HC, etc. The correlation between $C_{algo, real}$ and $C_{h, sub}$ is 0.950 and the correlation between $C_{algo, real}$ and $C_{h, fit}$ is 0.950. The correlation between $C_{h, sub}$ and $C_{h, fit}$ is 0.999. 
The conclusion from this last figure is that our metrics, either fitted or measured, are very well related with what a human would be experiencing in terms of complexity. This means that our model can be run on exotic scenarios in order to have a good idea of the workload that a human controller would experience. Note also that, with the results from \ref{hetero}, we can also conclude that the free-routing would in fact decrease the complexity of the airspace. This counter-intuitive result should be considered with caution, because essentially it resides in the following mechanism. On one hand, using free-routing decreases the number of heading and altitude changes, as well as the average distance between aircraft. On the other hand, the complexity of conflicts increase because of problematic geometries. As a results, the precise output depends on the real weights that the controller gives to these mechanisms.

In Table \ref{tab:OLS1} we provide the information about the weights with which the four PCA components enter the metric $C_{h, sub}$ described above. The same information is given in table \ref{tab:OLS2} for the metric $C_{h, fit}$. The comparison with Table \ref{tab:OLS} shows that indeed the role of the four PCA components is strikingly similar both for human controllers, see Table \ref{tab:OLS1} and our simulator, see Table \ref{tab:OLS}. This indicates that indeed our simulator  tackles complexity as much as humans in the current Air Traffic Management scenario.
\begin{table}[H]
\begin{center}
\begin{tabular}{c||c|c|c|c}
Comp.	& Weights	&	Std. err.	&	$P>|t|$	& 95\% CI  \\
\hline
\hline
0		& 0.2993	& 0.004			& 0.000		& [0.292, 0.306]\\
1		& 0.2440	& 0.008			& 0.000		& [0.228, 0.260]\\
2		& -0.0442	& 0.011			& 0.000		& [-0.065, -0.023]\\
3		& 0.0925	& 0.014			& 0.000		& [0.065 0.120]\\
\end{tabular}
\caption{Fit of $C_{h, sub}$ against the four components of the PCAs. $R^2 = 0.919$}  \label{tab:OLS1}
\end{center}
\end{table}

\begin{table}[H]
\begin{center}
\begin{tabular}{c||c|c|c|c}
Comp.	& Weights	&	Std. err.	&	$P>|t|$	& 95\% CI  \\
\hline
\hline
0		& 0.3015	& 0.003			& 0.000		& [0.295, 0.308]\\
1		& 0.2376	& 0.008			& 0.000		& [0.222, 0.253]\\
2		& -0.0416	& 0.010			& 0.000		& [-0.062, -0.021]\\
3		& 0.0896	& 0.013			& 0.000		& [0.063 0.116]\\
\end{tabular}
\caption{Fit of $C_{h, fit}$ against the four components of the PCAs. $R^2 = 0.925$}  \label{tab:OLS2}
\end{center}
\end{table}

\section{Conclusions} \label{concl}

In this paper we have presented an agent-based simulator able to simulate the air traffic management in the current and in the SESAR scenario. The model is composed of several modules fully described in \cite{git_repo_model_description}. The model simulates both the strategic phase associated to the planning of the aircraft trajectories and the tactical modifications that might occur in the en-route phase. It is able to forecast some high-level features of new scenarios, for instance when specific SESAR solutions are implemented.

Studying more specifically the free-routing solutions envisioned by SESAR, we have shown that in this scenario we can expect the controllers to perform a smaller number of operations although they will be dispersed over a larger portion of the airspace. This would in principle indicate an increase in the complexity controllers will have to deal with. In order to investigate this specific aspect, we have considered some metrics that are widely used in the literature to ``measure'' complexity in a certain airspace. These metrics are not independent from each other, as they measure complexity from different perspectives. Hence, we have used PCA to decrease the number of relevant metrics while keeping most of their predicting power. After having selected four components, which have quite physical interpretations, we tested if they have also the power of predicting the variation of the complexity for our algorithm, i.e. the number of actions, and the subjective variations of complexity for humans, with linear regressions. Since the regressions are quite good, we conclude that 1) the reduction to four components is able to catch most of the complexity in the sector 2) that our algorithm that can be safely used in new scenarios in order to predict the workload for humans. This good capacity forecast could be used by policy maker to test free-routing in denser regions than it is today.



Our results have been obtained in the idealized case when no sector is present in the considered airspace. This would correspond to the most extreme case where a huge portion of airspace is completely integrated, at least at cruse altitude. It is left for a future work to investigate how the possible partitions of the airspace into sectors would impact the management of trajectories in the SESAR scenario and whether or not this would lead to an increase in the complexity when it comes to having different controllers.



\section*{Aknowledgements}
We thank Marc Bourgois for fruitful discussion.

{\bf{Disclaimer}} - {\em{This work is co-financed by EUROCONTROL acting on behalf of the SESAR Joint Undertaking (the SJU) and the EUROPEAN UNION as part of Work Package E in the SESAR Programme. Opinions expressed in this work reflect the authors' views only and EUROCONTROL and/or the SJU shall not be considered liable for them or for any use that may be made of the information contained herein.}}

\appendix

\newpage 

\section{List of Complexity Metrics} \label{app1}

\begin{table} [H]
\caption{Complexity metrics used in section \ref{compmet} found in the literature. A quick description of each metric is included.}
\centering
\begin{tabular}{|c||c|c|}
\hline
\textbf{Metrics} &    \textbf{Based on} 													&   \textbf{Ref.}\\
\hline
\hline
C1     & Density 																			& \cite{Chatterji2001}\\
\hline
C2     & Number of climbing aircraft 														& \cite{Chatterji2001}\\
\hline
C3     & Number of stationary aircraft 													& \cite{Chatterji2001}\\
\hline
C4     & Number of descending aircraft 													& \cite{Chatterji2001}\\
\hline
C5     & Inverse of the mean weighted horizontal distance between aircraft pairs 			& \cite{Chatterji2001}\\
\hline
C6     & Inverse of the mean weighted vertical distance between aircraft pairs			& \cite{Chatterji2001}\\
\hline
C7     & 
		\begin{tabular}{c}
			Mean inverse horizontal distance of aircraft below vertical\\
			 separation distance
		\end{tabular}
		 																					& \cite{Chatterji2001}\\
\hline
C8     & 
		\begin{tabular}{c}
			Mean inverse vertical distance of aircraft below horizontal\\
			 separation distance
		\end{tabular}
																						 	&\cite{Chatterji2001}\\
\hline
C9     & Inverse minimal horizontal distance among pairs of aircraft				 		& \cite{Chatterji2001}\\
\hline
C10    & Inverse minimal vertical distance among pairs of aircraft 						& \cite{Chatterji2001}\\
\hline
C11    & Number of aircraft pairs with positive time-to-conflict					 		&\cite{Chatterji2001}\\
\hline
C12    &   
		\begin{tabular}{c}
			Number of aircraft with at least with positive\\ 
			time-to-conflict divided by mean time-to-conflict
		\end{tabular} 																		&\cite{Chatterji2001}\\
\hline
C13    & Inverse of minimal time-to-conflict 												& \cite{Chatterji2001}\\
\hline
C16    & Mean geometrical complexity based on crossing angles							 	& \cite{Chatterji2001}\\
\hline
\hline
N      & Density																			&\cite{Laudeman1998,Sridhar1998} \\
\hline
HC     & Number of aircraft with heading change greater than 15$^{\circ}$					& \cite{Laudeman1998,Sridhar1998}\\
\hline
AC     & Number of aircraft with altitude change greater than 750 feet					& \cite{Laudeman1998,Sridhar1998}\\
\hline
MD5    & Number of aircraft with 3d distance between 0-5 nautical miles 					& \cite{Laudeman1998,Sridhar1998}\\
\hline
MD10   & Number of aircraft with 3d distance between 6-10 nautical miles 					& \cite{Laudeman1998,Sridhar1998}\\
\hline
CP25   & 
		\begin{tabular}{c}
			Number of aircraft with horizontal distance between 0-25 nautical\\
			miles and vertical separation less than 2000 feet.
		\end{tabular}																		& \cite{Laudeman1998,Sridhar1998}\\
\hline
CP40   &   
		\begin{tabular}{c}
			Number of aircraft with horizontal distance between 26-40 nautical\\
			miles and vertical separation less than 2000 feet.
		\end{tabular}																		
																							&\cite{Laudeman1998,Sridhar1998}\\
\hline
CP75   & 
		\begin{tabular}{c}
			Number of aircraft with horizontal distance between 41-70 nautical\\
			miles and vertical separation less than 2000 feet.
		\end{tabular} 	
																							& \cite{Laudeman1998,Sridhar1998}\\
\hline
\end{tabular}
\end{table}

\newpage 

\section{Complexity Metrics in the SESAR scenario} \label{annex:pca_SESAR}

In section \ref{compmet} we clarified that our simulator works in a way which is quite close to that of a human controller in the current ATM scenario. We can therefore be confident that the results of section \ref{eff} and \ref{hetero}, based on numerical simulations relative to the SESAR scenario, are a reasonable forecast of how a human controller would behave in the SESAR scenario.

\begin{figure}[H]
\begin{center}
                       \includegraphics[width=0.45\textwidth]{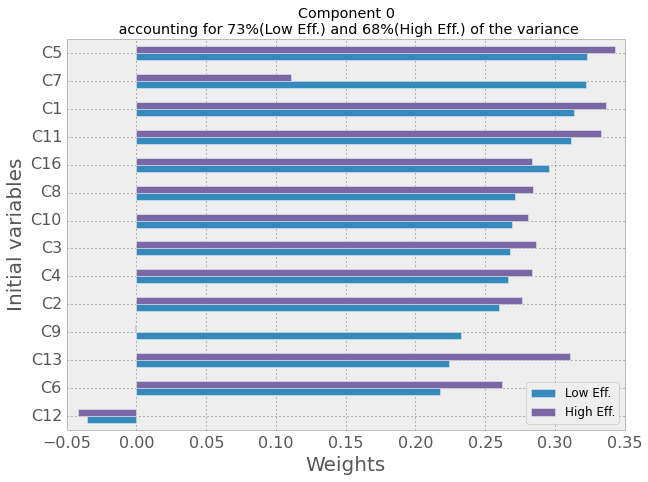}
                       \includegraphics[width=0.45\textwidth]{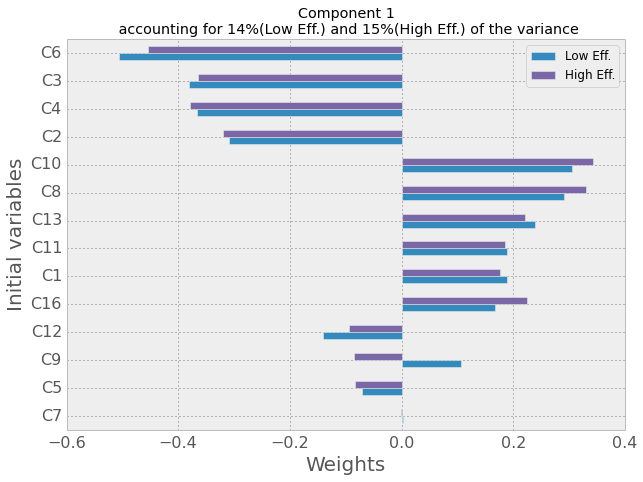}
                       \includegraphics[width=0.45\textwidth]{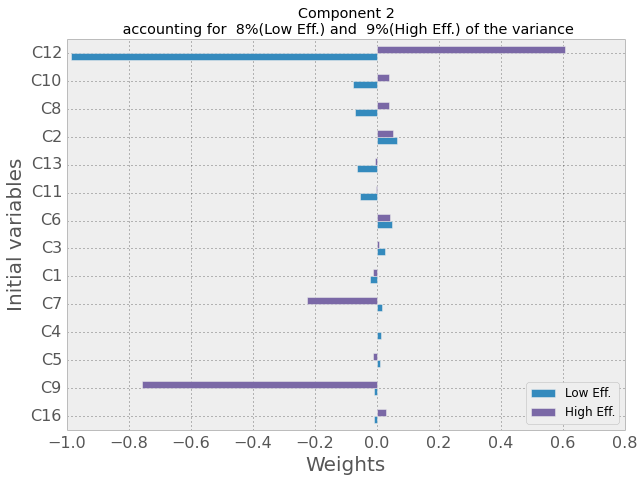}
                       \includegraphics[width=0.45\textwidth]{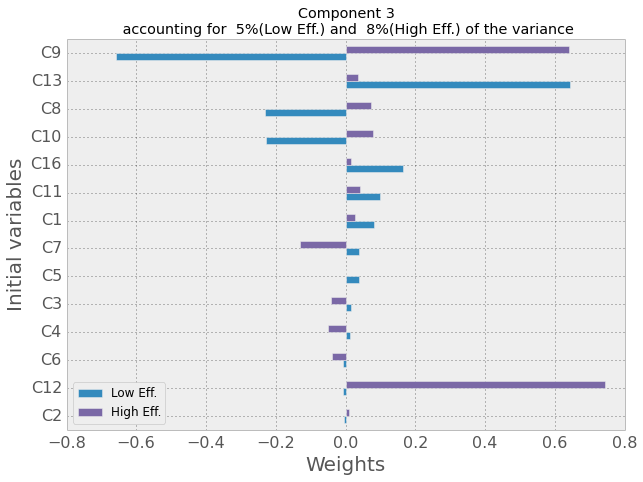}
\end{center}
\caption{Composition of the first four eigenvectors of the PCA in the SESAR scenario.} \label{fig:pcaSESAR}
\end{figure}

However, one might go further and check whether or not the determinants of complexity will be the same in the current and the SESAR scenario. Specifically, one might ask whether the four eigenvectors of Fig. \ref{fig:pca} will be the same also in the SESAR scenario. To this end, here we will consider numerical simulations obtained starting from deconflicted trajectories with highest efficiency E=0.999. As much as in the current scenario case, in Fig. \ref{fig:pcaSESAR} we show the composition, in terms of the initial variables C1 -- C16, of the four PCA eigenvectors with largest eigenvalues (violet). For the sake of comparison, we also show the composition for the current scenario (blue), i.e. the same data showed in Fig. \ref{fig:pca}. In the SESAR scenario these eigenvectors account for about 82\% of  the variance. The figure shows that indeed the composition of the four PCA eigenvectors is different in the current and SESAR scenario. In particular, the role of the C9 metrics seems to be drastically different in the two cases, while other components, such as C12 and C13 show differences in limited cases.

Again, our next step is to discover how well these eigenvectors can explain the variation in the number of actions of our SESAR scenario controllers. In order to do this, we perform and Ordinary Least Square regression on the data \cite{J.N.Miller2005}. The explicatory variables are the four components isolated above and the variable to explain is the number of action (na\_{tot}) of our controller. The result of the fit is presented in Table \ref{tab:OLSsesar}. A direct comparison with Table \ref{tab:OLS} shows that the weights are quite close to each other, thus confirming that the role of these four eigenvectors in explaining the complexity in the SESAR scenario is essentially the same as in the current scenario.
\begin{table}[H]
\begin{center}
\begin{tabular}{c||c|c|c|c}
Comp.	& Weights	&	Std. err.	&	$P>|t|$	& 95\% CI  \\
\hline
\hline
0        &       0.2640    &        0.006     &         0.000        &         [0.252     0.276]       \\
1        &       0.2571    &        0.014     &         0.000        &         [0.229     0.285]       \\
2        &      -0.0555    &        0.019     &         0.003        &        [-0.093    -0.018]       \\
3        &       0.0766    &        0.025     &         0.002        &         [0.028     0.125]       \\
\end{tabular}
\caption{Fit for $C_{h, fit}$. Ordinary Least Square method used to make a regression of the number of actions in our model against the four components found during the PCA. The first column present the weights, the second one the standard error, the third one a t-test for the coefficient, and the last one the 95\% confidence interval. Note that the variables have been standardized (i.e. we subtract the mean and divide by the standard deviation). The fit has a $R^2$ of 0.925.}  \label{tab:OLSsesar}
\end{center}
\end{table}

These interesting considerations will be studied in a subsequent paper.

\newpage
\bibliography{main_bibliography,additional_bib}{}

\end{document}